\documentclass[reprint,aps,prx,superscriptaddress,nofootinbib,longbibliography]{revtex4-2}
\usepackage[utf8]{inputenc}
\usepackage[american,british]{babel}
\usepackage[T1]{fontenc}
\usepackage{geometry}
\usepackage{amsmath,amsthm,amssymb,mathrsfs,mathtools,amsfonts}
\usepackage[pdftex]{graphicx}
\usepackage{xcolor}
\usepackage{dcolumn}
\usepackage{verbatim}
\usepackage{epsfig}
\usepackage{color}
\usepackage{xfrac}
\usepackage{bbold}
\usepackage{braket}
\usepackage{tensor}
\usepackage[normalem]{ulem}
\usepackage{bm}
\usepackage{subfigure}
\usepackage{float}
\usepackage{booktabs}
\usepackage{slashed}
\usepackage{tikz}
\usepackage{cancel}
\usepackage{wasysym}
\usepackage[bookmarks=false]{hyperref}
\hypersetup{
    pdfstartview={FitH},
    pdftitle={Interface phases and dynamics in two-dimensional quantum magnets: A holographic approach from universality to quantum simulation},
    pdfauthor={Abhishodh Prakash, Jaydev Singh Rao, S. A. Parameswaran, Alessio Lerose},
    pdfnewwindow=true,
    colorlinks=true,
    linkcolor=blue,
    citecolor=blue,
    filecolor=magenta,
    urlcolor=blue
}
\usepackage[capitalize]{cleveref}
\usepackage{orcidlink}

\renewcommand{\sim}{\thicksim}

\newcommand{\A}{{\cal A}}
\newcommand{\B}{{\cal B}}
\newcommand{\C}{{\cal C}}

\begin{document}

\newcommand{\titleinfo}{
Interface phases and  dynamics in two-dimensional quantum magnets: \\
A ``holographic'' approach from universality to quantum simulation
}

\title{\titleinfo}

\author{Abhishodh Prakash~\orcidlink{0000-0001-8096-8596}}
\thanks{These authors contributed equally.}
\affiliation{The
Rudolf Peierls Centre for Theoretical Physics, Oxford University, Oxford OX1 3NP, United Kingdom
}
\affiliation{Harish-Chandra Research Institute (HRI), Prayagraj (Allahabad) 211019, India}
\affiliation{Homi Bhabha National Institute (HBNI), Mumbai 400094, India}

\author{Jaydev Singh Rao~
\orcidlink{0009-0005-1994-6933}}
\thanks{These authors contributed equally.}
\affiliation{Institute for Theoretical Physics, KU Leuven, Celestijnenlaan 200D, 3001 Leuven, Belgium}

\author{Siddharth A. Parameswaran~\orcidlink{0000-0002-5055-5528}}

\affiliation{The
Rudolf Peierls Centre for Theoretical Physics, Oxford University, Oxford OX1 3NP, United Kingdom
}

\author{Alessio Lerose~
\orcidlink{0000-0003-1555-5327}}
\affiliation{The
Rudolf Peierls Centre for Theoretical Physics, Oxford University, Oxford OX1 3NP, United Kingdom
}
\affiliation{Institute for Theoretical Physics, KU Leuven, Celestijnenlaan 200D, 3001 Leuven, Belgium}

\begin{abstract}
We introduce a framework to classify quantum phases, phase transitions, and non-equilibrium dynamics of interfaces separating ordered bulk
domains in 2D quantum magnets --- equivalently, confining strings in dual lattice gauge theories ---
based on effective 1D Hamiltonians governing geometric fluctuations.
Building on a ``holographic'' approach from \mbox{[Phys. Rev. Lett. \textbf{129}, 120601 (2022)]}, here reinterpreted as an \textit{exact bosonization},
we uncover a rich
quantum phase structure, with a  variety of stiff and rough  interface phases described by gapped and gapless 1D ground states, \mbox{respectively, all} distinguishable through the statistics of 2D wave-function snapshots.
Our framework allows us to predict distinct spatiotemporal scaling laws for non-equilibrium curvature-driven interface dynamics   across parameter space,
which can be readily probed in existing experiments.
We finally show that our approach enables the unprecedented experimental opportunity of \textit{directly} measuring charge full counting statistics and symmetry-resolved properties of an encoded 1D system, as we explicitly demonstrate by numerically simulating a neutral-atom array experiment.
\end{abstract}

\date{\today}
\maketitle

\tableofcontents
\section{Introduction}

Through pioneering experimental advances, the precision study of entangled states and far-from-equilibrium dynamics in two-dimensional (2D) quantum matter has transformed a long-standing theoretical frontier into an established field of experimental quantum science.
Driven by challenges in quantum simulation of fundamental physics and in the implementation of  quantum-computing primitives~\cite{terhal2015quantum,briegel2009measurement}, several platforms such as neutral-atom arrays~\cite{ebadi2021quantum,semeghini2021probing,scholl_quantum_2021,leclerc2026one,bornet2026dirac}, superconducting qubits~\cite{kim2023evidence,andersen2024thermalization,cochran2025visualizing}, trapped ions~\cite{kiesenhofer2023controlling,guo2024site,haghshenas2026digital}, and optical lattices~\cite{sun2021realization,adler2024observation,karch2026dynamical} {have} emerged as simulators of effective 2D quantum magnets, with the added feature that their geometry and interactions are both highly configurable. These developments have opened new avenues in quantum simulation, and early explorations have already led to the observation of signatures of various kinds of phenomena such as magnetic ordering~\cite{scholl_quantum_2021,ebadi2021quantum,chen2023continuous,leclerc2026one}, spin-liquidity~\cite{semeghini2021probing,karch2026dynamical,bornet2026dirac}, confining strings~\cite{cochran2025visualizing,gonzalez2025observation,joshi2026observation}, fractonic excitations~\cite{adler2024observation}, and quantum coarsening dynamics~\cite{andersen2024thermalization,manovitz2025quantum}.

In this work, we study quantum phases, phase transitions and non-equilibrium dynamics of {system-wide topological defect lines}, i.e., magnetic interfaces, in 2D synthetic quantum spin lattices. In classical statistical mechanics, it has long been known
that extended interfaces can exhibit distinct large-scale fluctuation behaviors. As a paradigmatic example, in the standard 3D Ising model, an interface separating positively and negatively magnetized regions exhibits a ``flat'' (or ``stiff'') phase at temperatures $0<T<T_r\approx 0.56 T_c$ with weak size-independent fluctuations around a lattice plane, and a ``rough'' phase with divergent fluctuations for $T_r<T<T_c$, where $T_c$ is the critical temperature for bulk ferromagnetism~\cite{Weeks1980,burkner1983monte,hasenbusch1997computing}. Statistical interface fluctuations can be captured by approximate dimensionally-reduced 2D geometric models~\cite{vanbeijeren1977exactly,frohlich1981kosterlitz,DIJKGRAAF2009463}, equivalently  by a 1D quantum transfer matrix. This approach has been successful in describing a variety of systems, from various behaviors of crystal-vapor surfaces~\cite{dennijs1989preroughening,BERNASCONI1993363,santoro1994disordered,carlon1997transfer} to confining strings in Euclidean lattice gauge theories~\cite{luscher1981thick,kogut_analyticity_1981,kogut_fluctuating_1981,CASELLE1996397} (which are related to magnetic interfaces by exact dualities~\cite{wegner1971}). Analogous effective theoretical descriptions have been proposed for quantum crystal surfaces~\cite{fradkin1983roughening,IordanskiiKorshunov1984,balibar1994helium,soyler2007superfluid} and stripes in high-$T_c$ superconductors~\cite{eskes1996quantizing}.

A common thread running through all these studies is that the interfaces possess a classical-statistical formulation, making their equilibrium phase diagrams efficiently accessible through Monte Carlo sampling techniques.  Recent works have started developing field-theoretical~\cite{kampanis2026interface} and tensor-network computational methods for ground states and real-time dynamics of quantum interfaces or equivalent confining strings~\cite{xu2025tensor,krinitsin_roughening_2025,Krinitsin2025time,marcantonio_roughening_2025,ueda_interface_2026,xu2025string}. However, the phase structure of general interfaces beyond the (relatively rare) sign-problem-free instances, as well as their universal non-equilibrium relaxation behaviors, appear to be largely unexplored.

In this work, we address these general questions, directly motivated by the possibilities
offered by modern quantum-simulation platforms and inspired by the recent pioneering experiment by Manovitz et al.~\cite{manovitz2025quantum}. Specifically, we ask:
\begin{enumerate}
    \item[Q1:] What is the landscape of quantum phases and phase transitions that can be induced in extended 1D topological defects  by  leveraging the
    powerful programmability of bulk geometry and Hamiltonian interactions available in 2D synthetic quantum spin arrays?
    \item[Q2:] What is the nature of real-time non-equilibrium quantum dynamics of these extended topological defects? What universal geometry-driven relaxation laws can emerge?
    \item[Q3:] Does the simulation of quantum interfaces create fundamentally new experimental opportunities compared to the conventional quantum-simulation toolbox?
\end{enumerate}
Our study provides the following answers:
\begin{enumerate}
    \item[A1:] We show that variations of the parent 2D magnetic Hamiltonian  generate an extremely rich phase structure --- essentially including the \textit{entire known phase diagram of $U(1)$-symmetric 1D quantum systems}~\cite{Giamarchi};
    \item[A2:] Curvature-driven quantum  dynamics can exhibit various distinct universal scaling behaviors, including  diffusive (corresponding, as we recognize, to the experimental findings of Ref.~\cite{manovitz2025quantum}), super-diffusive, and ballistic scaling laws;
    \item[A3:] Viewed as a simulator of an effective lower-dimensional quantum system, the setup offers unique experimental advantages, such as the possibility of {directly} manipulating and measuring  \textit{global} observables via \textit{local} operations.
\end{enumerate}

On a technical level, we leverage a ``holographic'' approach that may be viewed as a generalization of the traditional dimensional-reduction approaches reviewed above. Recently, this approach was introduced in a form close to ours by Balducci et al. in Refs.~\cite{Alessio_PhysRevLett.129.120601,Alessio_PhysRevB.107.024306} (including one of us) to study the real-time evolution of false-vacuum bubbles in the 2D quantum Ising model; previously, a similar construction was introduced by Kogut et al. to study the restoration of rotational symmetry in the continuum limit of confining lattice gauge theories~\cite{kogut_fluctuating_1981}. Building on these ideas, in this work, we construct a general theoretical framework and  propose a key interpretation for it as an \textit{exact bosonization} at lattice level,  fully clarifying its relationship to field-theoretical bosonization~\cite{Giamarchi}.

Although in this work we will mostly use the language of interfaces in 2D quantum magnets, as anticipated above, all our results and discussion can be recast in terms of the dual picture of confining strings in lattice gauge theories~\cite{drouffe1981roughening1,kogut_fluctuating_1981}. Our choice is mainly motivated by the straightforward connection between quantum magnets and the native Hamiltonians describing several quantum-simulation platforms.

\begin{figure*}
\begin{subfigure}
\centering
\includegraphics[width=0.28\textwidth]{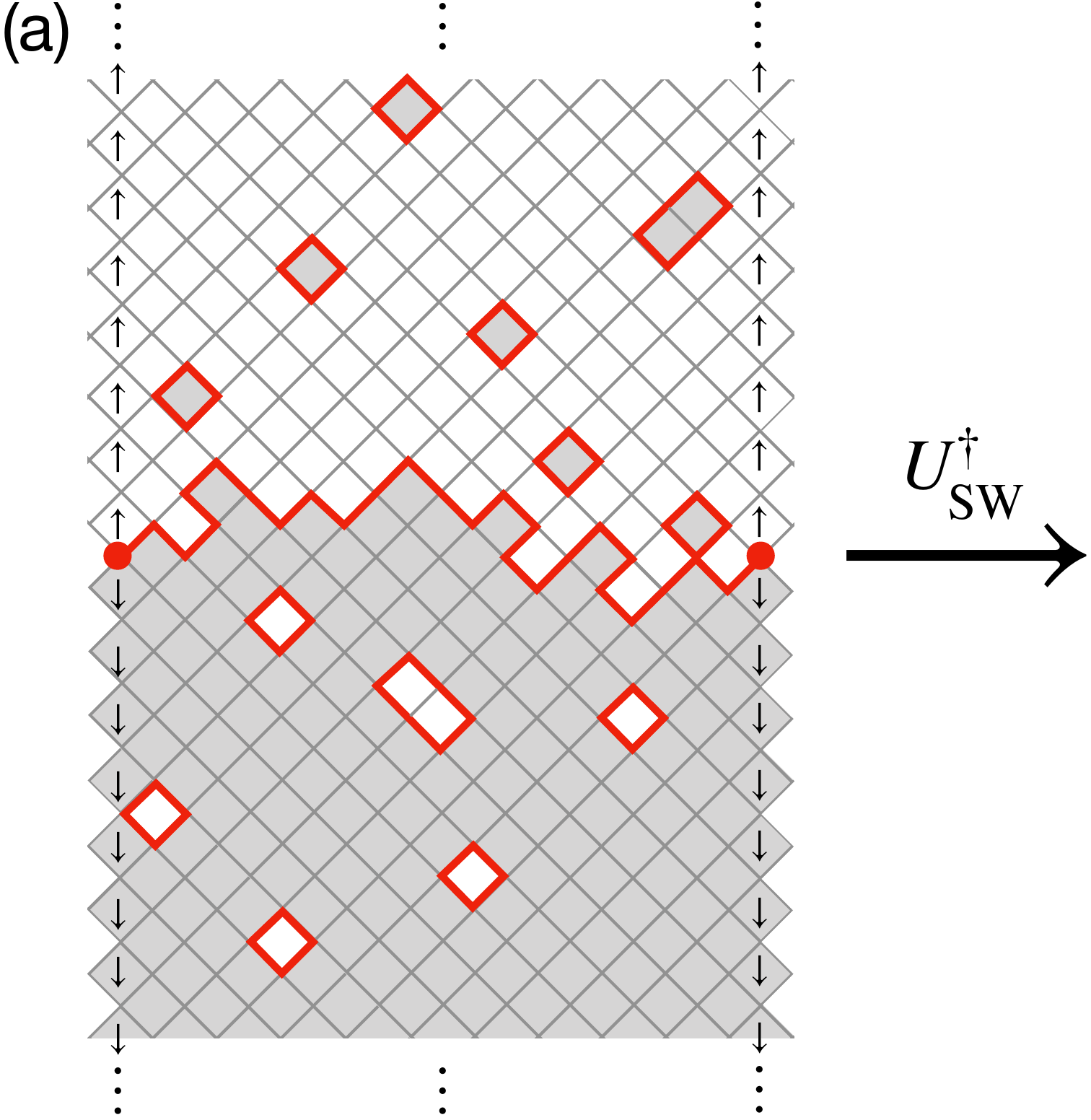}
\end{subfigure}
\begin{subfigure}
\centering
\includegraphics[width=0.28\textwidth]{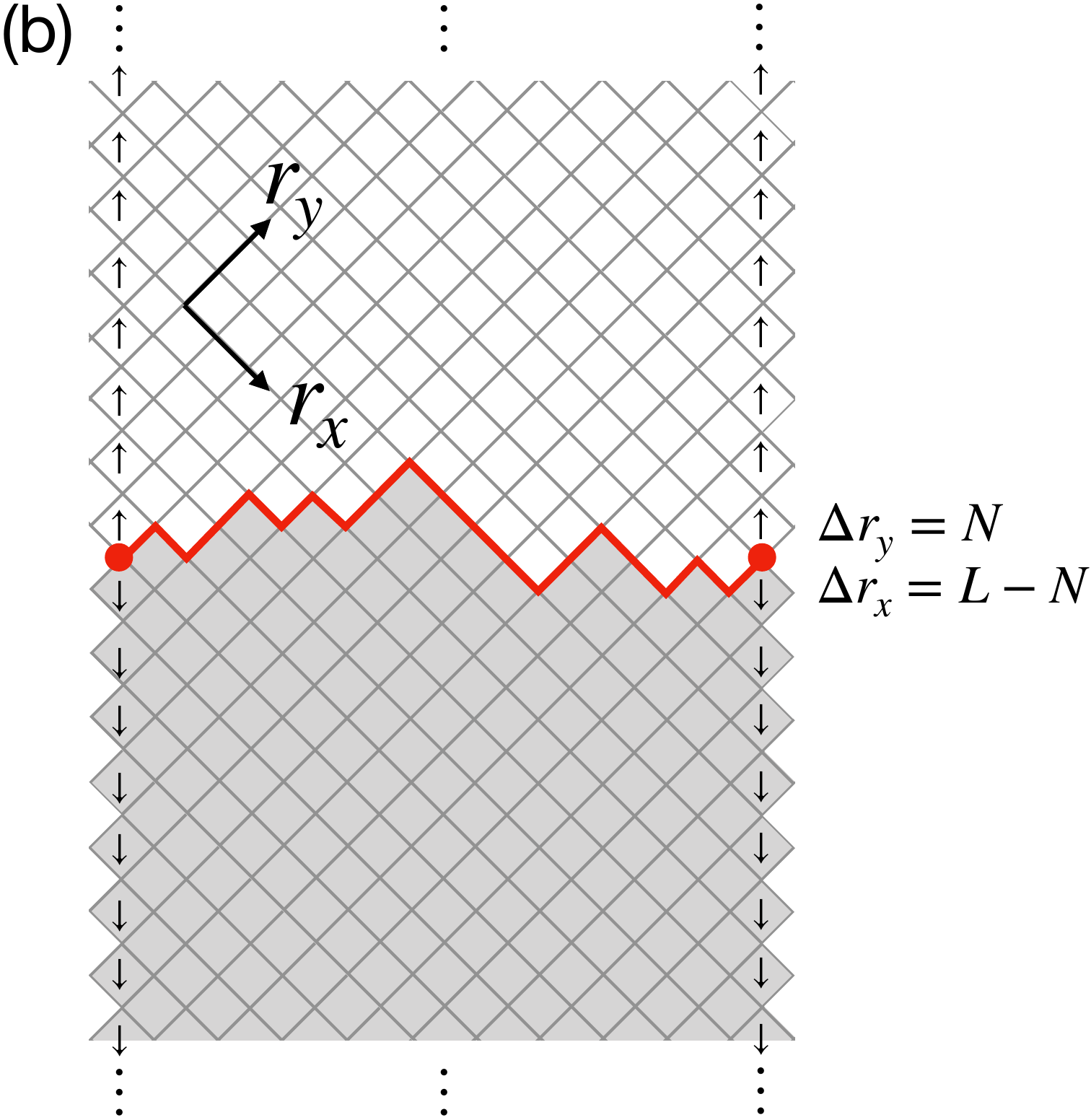}
\end{subfigure}
\begin{subfigure}
\centering
\raisebox{4mm}{
\includegraphics[width=0.35\textwidth]{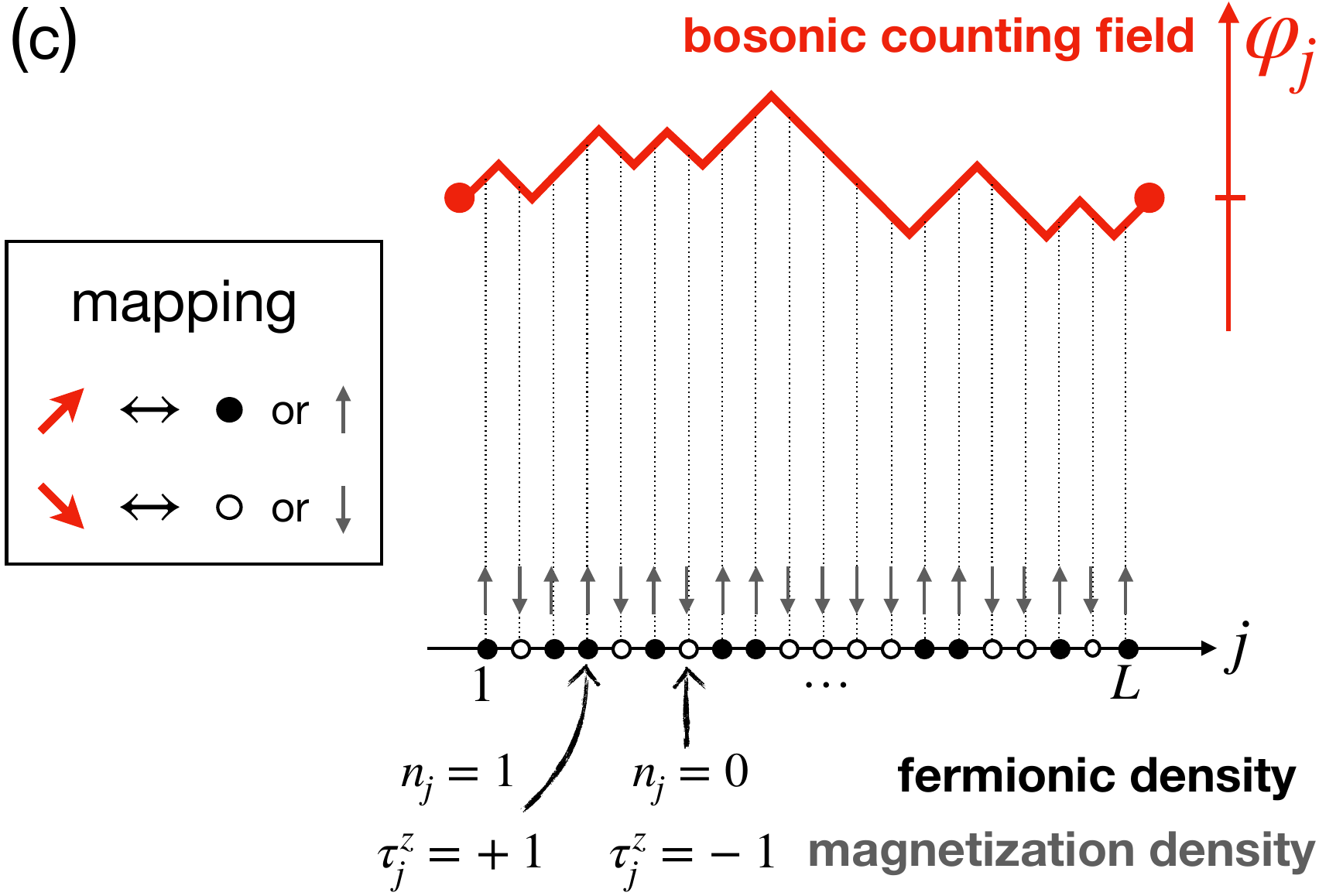}
}
\end{subfigure}
\caption{
\textbf{``Holographic'' description of interfaces in 2D quantum magnets.}
For illustration, we consider an Ising square lattice tilted by $45^\circ$. Spins on the vertical sides are frozen as shown, pinning the endpoints of a domain wall.
Panel (a):
Typical computational-basis snapshot of the ground-state wave function of a quantum Ising-type ferromagnet with the considered boundary conditions, with white/gray  plaquettes depicting up/down spins.
Panel (b): Undoing the Schrieffer-Wolff transformation $U_{\rm SW}$, the bulk regions look perfectly magnetized, and wave function snapshots consist of minimal-length domain walls, i.e., ground states of the classical Ising ferromagnet.
Panel (c): The classical ground-state manifold is ``holographically'' mapped to 1D Fock space of a fermionic [or spin-$1/2$] chain as shown. The fermionic filling $\rho=N/L$ [magnetization density $m=M/L=2\rho-1$] is fixed and geometrically determined by the  orientation angle~$\Theta$ formed by the pinned domain-wall endpoints via $\tan \Theta =  m$. These ``holographic'' 2D configurations are equivalently described by the height field $\varphi_j=2\sum_{l=1}^j c^\dagger_l c_l - j$ [$=\sum_{l=1}^j \tau^z_l$].
}
\label{fig_mapping}
\end{figure*}

\section{``Holographic'' description of interfaces}

\label{sec_mapping}

To illustrate the general ideas of this work, we first consider the simplest possible setup, with spin-$\frac{1}{2}$
 particles occupying the sites of a square lattice with nearest-neighbor Ising couplings of strength $J$. The system is described by the Hamiltonian
\begin{equation}
\label{eq_classicalIsing}
    H_{\rm 2D}^{(0)} = - J \sum_{\mathbf{r}} \left( \sigma^z_{\mathbf{r}}\sigma^z_{\mathbf{r}+\mathbf{e}_x}+\sigma^z_{\mathbf{r}}\sigma^z_{\mathbf{r}+\mathbf{e}_y}\right),
\end{equation}
where $\mathbf{r}=(r_x,r_y)$ are integer lattice coordinates, $\mathbf{e}_x=(1,0)$ and $\mathbf{e}_y=(0,1)$ are the lattice unit vectors, and $\sigma^{\mu}_{\mathbf{r}}$, $\mu=x,y,z$ are Pauli matrices with support on site $\mathbf{r}$.
For our purposes, it will be convenient to work with a rectangular array rotated by $45^\circ$ with respect to the lattice axes, and with boundary spins frozen in a configuration that enforces a domain wall across the system, as illustrated in Fig.~\ref{fig_mapping}(a,b).

\subsection{Classical low-energy configurations}

\label{sec_configs}

With the considered boundary conditions, the classical ground states of Eq.~\eqref{eq_classicalIsing} form a highly degenerate set associated with the spin configurations with shortest domain walls separating perfectly magnetized regions.
With reference to Fig.~\ref{fig_mapping}(b), if the domain-wall endpoints have relative coordinates $\Delta r_x = N$ and $\Delta r_y = L-N$, with $0\le N \le L$, one can check that the minimal-length domain walls are in one-to-one correspondence with the set of directed paths traveling eastwards at each step, with exactly $N$ north-east and $L-N$ south-east steps. Each such directed path draws a height function $\varphi_j$ with respect to some horizontal baseline, as shown by the red line in Fig.~\ref{fig_mapping}(c).

The characterization above shows that the ground-state manifold of the Hamiltonian~\eqref{eq_classicalIsing} in the considered geometry can be ``holographically'' mapped to the $\binom{L}{N}$-dimensional Hilbert space of $N$ fermionic particles in a 1D chain of length $L$~\cite{Alessio_PhysRevLett.129.120601,Alessio_PhysRevB.107.024306}, as illustrated in Fig.~\ref{fig_mapping}(c).
Identifying fermions with $\uparrow$-spins and holes with $\downarrow$-spins, this Hilbert space can be equivalently described in terms of a spin-$\frac12$ chain of length $L$ with fixed total magnetization $M=2N-L$.
In this representation, the local fermionic occupation number $c^\dagger_j c_j=0,1$ or spin magnetization $\tau^z_j=-1,+1$ corresponds to the local path slope at the abscissa $j$ being $-1$ or $+1$, respectively.
In turn, the height function $\varphi_j=2\sum_{l=1}^j c^\dagger_l c_l-j=\sum_{l=1}^j \tau^z_l$ is ``holographically'' interpreted as a bosonic counting field for the 1D fermions or spins.
The filling factor $\rho=N/L$ or the magnetization density $m=M/L=2\rho-1$ is geometrically related to the slope that connects the fixed endpoints of the domain wall, $\tan \Theta = \frac{\varphi_L-\varphi_0}{L} = m$.

\subsection{Quantum low-energy dynamics}

\label{sec_SW}

We now consider finite-range classical or quantum perturbations $V_{{\rm 2D}, \alpha}$, $\alpha=1,2,\dots$, to the Ising Hamiltonian~\eqref{eq_classicalIsing}
that preserves its $\mathbb{Z}_2$ symmetry. The strengths $\lambda_\alpha$ of these perturbations are collected in an array $\vec{\lambda}$. When the overall energy scale $\lambda=|\vec{\lambda}|$ associated with $V_{\rm 2D}=\sum_\alpha V_{{\rm 2D}, \alpha} $ is much smaller than the gap $\lambda\ll 8J$, the bulk ferromagnetic ground states are weakly dressed by dilute magnetization-defects while retaining their spontaneous-symmetry-breaking nature. In particular, the domains retain their stability. On the other hand, the perturbation has a dramatic effect on the interface: a quantum perturbation of  infinitesimal strength $\lambda\neq0$ lifts the exponential ground-state degeneracy  of classical domain-wall configurations, selecting appropriate quantum superpositions as ground states. For small but finite perturbations, this ``holographic'' 1D wave function is weakly dressed by dilute interface defects such as overhangs, analogously to the bulk, as shown in Fig.~\ref{fig_mapping}(a).

Our formal tool to map out interface quantum ground states and dynamics is the Schrieffer-Wolff transformation~\cite{McDonaldSchiefferWolffHubbard,FrohlichSchriefferWolff,LossSchriefferWolff}. This well-established method allows us to obtain an effective Hamiltonian $H_{\rm SW}=H_{\rm 2D}^{(0)} + \lambda H_{\rm 2D}^{(1)} + \frac {\lambda^2} J H_{\rm 2D}^{(2)} + \dots $
that preserves the degenerate eigenspaces of
$H_{\rm 2D}^{(0)}$, through a sequence of unitary transformations
$U_{\rm SW}=e^{i \frac \lambda J S^{(1)} }e^{i \frac {\lambda^2} {J^2} S^{(2)} } \cdots$, i.e., \begin{equation}
    U^\dagger_{\rm SW} \left( H_{\rm 2D}^{(0)}+ V_{\rm 2D} \right) U_{\rm SW} = H_{\rm SW}
    \end{equation}
order-by-order in $\lambda/J$.
The effective Hamiltonian interactions $H_{\rm 2D}^{(n)}$ and the hermitian Schrieffer-Wolff generators $S^{(n)}$ are calculated recursively in~$n$.
By construction, $H_{\rm SW}$ is block-diagonal, i.e., it
preserves subspaces with fixed total domain-wall length (including the classical ground-state manifold), while $U_{\rm SW}$ describes the rotation of these subspaces in the full Hilbert space, which we refer to as dressing.

In quantum many-body systems, the Schrieffer-Wolff formal perturbative scheme for $H_{\rm SW}$ and $U_{\rm SW}$ does not converge to well-defined operators. Rather, in general, the scheme can be truncated to a finite optimal order, and the resulting truncated $H_{\rm SW}$ and $U_{\rm SW}$ describe \textit{prethermal} quantum dynamics over a  timescale that diverges exponentially with $J/\lambda$~\cite{abanin2017rigorous,lin2017quasiparticle,lerose2020quasilocalized}, as we will further briefly review in Sec.~\ref{sec_neq}. Projection of $H_{\rm SW}$ and $U_{\rm SW}$ to the ground-state manifold, however, is thought to be equivalent to standard cluster expansions and yield, in many cases, a finite radius of convergence. In this work, however, we will not be concerned with convergence issues, and we will treat truncations of $H_{\rm SW}$ and $U_{\rm SW}$ to suitably large finite orders  as accurate descriptions of static and dynamical interface behavior for small but finite perturbation strength.\footnote{Our numerical calculations of low-energy states in fairly large 2D spin lattices in Sec.~\ref{sec_qusim} quantitatively confirm the validity of a description of the 2D ground state based on $H_{\rm SW}$ and $U_{\rm SW}$. }

The two dressed bulk ground states of $H_{\rm 2D}=H_{\rm 2D}^{(0)}+ V_{\rm 2D}$ are given by $|\Psi_{\uparrow,\downarrow}\rangle = U_{\rm SW} |\Psi^{(0)}_{\uparrow,\downarrow}\rangle$, where the ground states $|\Psi^{(0)}_{\uparrow,\downarrow}\rangle$ of $H_{\rm SW}$ are the two perfectly magnetized classical configurations. For small $\lambda/J$, the dressed wave functions can be pictured as dilute superpositions of defects (i.e., flipped spins) with density $\mathcal{O}\left((\lambda/J)^2\right)$, generated by $e^{i\lambda/J S^{(1)}}$, cf. Fig.~\ref{fig_mapping}(a).\footnote{The quantum tunneling amplitude $\langle \Psi^{(0)}_\uparrow | H_{\rm SW} | \Psi^{(0)}_\downarrow \rangle$ is exponentially suppressed in system size.}

As discussed in Sec.~\ref{sec_configs} and illustrated in Fig.~\ref{fig_mapping}, with our choice of boundary conditions, the classical ground-state manifold is given by the ``holographic'' Hilbert space mapped to a 1D fermionic or spin-chain Fock space.
Denoting as $P_{\rm 1D}$ the projection operator onto this subspace, we obtain an effective 1D Hamiltonian $H_{\rm 1D} = P_{\rm 1D} H_{\rm SW} P_{\rm 1D}$ governing low-energy dynamics of the system to all orders in $\lambda/J$.
At the lowest orders, we have
\begin{equation}
\label{eq_HSW}
\begin{split}
    H_{\rm 1D}
    =   \sum_\alpha & \lambda_\alpha P_{\rm 1D} V_{{\rm 2D},\alpha} P_{\rm 1D} \\
    & + \lambda_\alpha^2 P_{\rm 1D} V_{{\rm 2D},\alpha} \frac{  \mathbb{1}-P_{\rm 1D}}{E_{\rm 1D}^{(0)}-H_{\rm 2D}^{(0)}} V_{{\rm 2D},\alpha} P_{\rm 1D}
    \\&+
    \dots \, ;
    \end{split}
\end{equation}
analogous, more cumbersome expressions provide higher-order corrections.

We will calculate the effective 1D Hamiltonian $H_{\rm 1D}$ and theoretically analyze and simulate its ground state  $|\Phi_{\rm 1D}\rangle$ --- or non-equilibrium dynamics $|\Phi_{\rm 1D}(t)\rangle$ --- in 1D Fock space [cf.~Fig.~\ref{fig_mapping}(c)] as a function of the parameters $\vec\lambda$ and $J$. Denoting as $|\Phi_{SW}\rangle$ the corresponding quantum superposition of classical minimal-length domain-wall configurations of the 2D spin lattice [cf.~Fig.~\ref{fig_mapping}(b)], obtained through the 1D-2D ``holographic'' identification, the dressed ground state $|\Phi_{\rm 2D}\rangle$ of $H_{\rm 2D}$ with our boundary conditions is given by $|\Phi_{\rm 2D}\rangle = U_{\rm SW} |\Phi_{{\rm SW}}\rangle$ [cf.~Fig.~\ref{fig_mapping}(a)]. This dressing needs to be kept in mind for experimental reconstruction of ``holographic'' predictions, as we will discuss later.

\subsection{Effective 1D quantum Hamiltonians}

\label{sec_H1D}

We will now derive the effective Hamiltonian $H_{\rm 1D}=\sum_\alpha H_{{\rm 1D},\alpha}$ governing interface quantum fluctuations for a range of possible bulk interactions $V_{{\rm 2D},\alpha}$ describing a 2D quantum magnet. $H_{\rm 1D}$ may be expressed in terms of equivalent degrees of freedom:
\begin{enumerate}
    \item[\textit{i)}] The height field~$\varphi$,
    \item[\textit{ii)}] $N$ interacting fermions in a chain of length~$L$,
    \item[\textit{iii)}] A spin chain with fixed magnetization $M$.
\end{enumerate}
In this Section, we will interchangeably speak about fermionic particles or a spin-$\frac{1}{2}$ chain,
the two representations being related via the standard Jordan-Wigner transformation; the representation in terms of $\varphi$ will be extensively discussed in Sec.~\ref{sec:noncompact_height}.
To avoid confusion with the original 2D spin operators $\sigma^{x,y,z}_{\mathbf{r}}$, we will use the notation $\tau^{x,y,z}_j$ for the Pauli matrices associated with the 1D spins describing the interface dynamics.
Crucially, our geometric setup in 2D imposes a strictly fixed number of particles or total magnetization, resulting in an \textit{exact $U(1)$ symmetry} of any induced effective 1D Hamiltonian.

The simplest symmetry-preserving perturbation to Eq.~\eqref{eq_classicalIsing} is a transverse magnetic field, $-g\sum_{\mathbf{r}}\sigma^x_{\mathbf{r}}$. The resulting lowest-order Schrieffer-Wolff Hamiltonian is a simple fermion hopping~\cite{Alessio_PhysRevLett.129.120601,Alessio_PhysRevB.107.024306}
\begin{equation}
\label{eq_1Dnnhopping}
    H_{\rm 1D} =
    -g \sum_{j=1}^{L-1} \left(c^\dagger_j c_{j+1} + {\rm H.c.}\right) \, ,
       \end{equation}
where a hop in 1D corresponds to flipping a domain-wall corner in 2D.
Let us now consider a next-nearest-neighbor Ising coupling, $- J^\prime\sum_{\mathbf{r}} \left( \sigma^z_{\mathbf{r}}\sigma^z_{\mathbf{r}+\mathbf{e}_x+\mathbf{e}_y}+\sigma^z_{\mathbf{r}}\sigma^z_{\mathbf{r}+\mathbf{e}_x-\mathbf{e}_y}\right)$. It is straightforward to check by direct inspection that this term generates an energy penalty for locally straight portions of the domain wall, equivalent to a density-density fermionic interaction:
\begin{equation}
\label{eq_1Ddensitydensity}
    H_{\rm 1D} =
     + 4J^\prime \,  \sum_{j=1}^{L-1} n_j {n_{j+1}}.
\end{equation}
Combining the perturbations leading to Eqs.~\eqref{eq_1Dnnhopping} and~\eqref{eq_1Ddensitydensity}, we obtain an interface described by an
interacting spinless-fermion chain,
equivalent to the XXZ quantum spin chain:
\begin{equation}
\label{eq_1DXXZ}
      H_{\rm XXZ} =  \sum_{j=1}^{L-1} -\frac g2 \left(\tau^x_j \tau^x_{j+1} + \tau^y_j \tau^y_{j+1} \right)
     + J^\prime \tau^z_j \tau^z_{j+1}.
     \end{equation}
(Notice the reversed sign in front of $J^\prime$ compared to $H_{\rm 2D}$.)
Intriguingly, interface dynamics acquires an emergent non-Abelian $SU(2)$ symmetry on fine-tuning $J^\prime=\pm g/2$, which has no bearing on bulk dynamics.\footnote{As is well known, the sign of $g$ is immaterial, as it can be flipped through a unitary transformation (e.g., a $\pi$-rotation about the $z$-axis of the even sublattice).}

Different bulk perturbations generate an even richer set of 1D interface Hamiltonians.
Consider, for example, a weak additional spin-spin interaction $-g^\prime_x\sum_{\mathbf{r}}\sigma^x_{\mathbf{r}}\sigma^x_{\mathbf{r}+\mathbf{e}_x}-g^{\prime}_y\sum_{\mathbf{r}}\sigma^x_{\mathbf{r}}\sigma^x_{\mathbf{r}+\mathbf{e}_y}$, with independently tunable couplings along the x and y lattice directions. The resulting lowest-order Schrieffer-Wolff Hamiltonian is $-g^\prime_x \sum_{j=1}^{L-2} (c^\dagger_j P_{\Circle_{j+1}}c_{j+2} + {\rm h.c.})+g^\prime_y \sum_{j=1}^{L-2} (c^\dagger_j P_{\CIRCLE_{j+1}}c_{j+2} + {\rm h.c.})$
where $P_{\Circle_{j}}$ and $P_{\CIRCLE_{j}}$ are local projection operators onto the empty and occupied fermionic state, respectively, at site~$j$. The sign reversal in the second term arises from the Jordan-Wigner transformation. Choosing $g^\prime_x=g^\prime_y \equiv g^\prime$ we obtain a longer-range XX spin chain. Together with longer-range diagonal Ising couplings --- which lead to longer-range ZZ interactions, see Sec.~\ref{sec_rydbergs} below --- we can obtain longer-range XXZ spin chains.
Choosing instead $g^\prime_x=-g^\prime_y \equiv g^\prime$, we obtain a next-nearest-neighbor fermion hopping,
\begin{equation}
\label{eq_1Dnnnhopping}
    H_{\rm 1D} =
    -g^\prime \sum_{j=1}^{L-2} \left(c^\dagger_j c_{j+2} + {\rm H.c.}\right)   \, .
\end{equation}

Let us now consider a next-nearest-neighbor Ising coupling with a checkerboard pattern,
$- \,\sum_{\mathbf{r}} J^\prime_\mathbf{r}\left( \sigma^z_{\mathbf{r}}\sigma^z_{\mathbf{r}+\mathbf{e}_x+\mathbf{e}_y}+\sigma^z_{\mathbf{r}}\sigma^z_{\mathbf{r}+\mathbf{e}_x-\mathbf{e}_y}\right) $ with $J^\prime_\mathbf{r}= J^\prime\left(1+(-)^{r_x+r_y}\delta\right)$. By generalizing the previous derivation for the uniform case, one can check that the induced fermionic interaction is
\begin{equation}
\label{eq_1Ddensitydensitystaggered}
    H_{\rm 1D} =
    +4J^\prime \sum_{j=1}^{L-1} \left(1+(-)^{j}\delta\right) n_j n_{j+1}    \, .
\end{equation}
Let us now combine the two perturbations leading to Eqs.~\eqref{eq_1Dnnnhopping} and~\eqref{eq_1Ddensitydensitystaggered}.
In this case, since the number of fermions occupying the even and odd subchains are independently conserved, we can treat them as two separate species and denote them by $\uparrow$ and $\downarrow$, i.e., $c^{(\dagger)}_{2l} \leftrightarrow c^{(\dagger)}_{l,\uparrow}$, $c^{(\dagger)}_{2l+1} \leftrightarrow c^{(\dagger)}_{l,\downarrow}$. On choosing $\delta=-1$ and $L$ even, we obtain the 1D (Fermi-)Hubbard Hamiltonian:
\begin{equation}
\label{eq_1DHubbard}
    H_{\rm FH} =
    -g^\prime
\sum_{\substack{l\\\sigma=\uparrow,\downarrow}}
(c^\dagger_{l,\sigma} c_{l+1,\sigma} + {\rm H.c.}) + 8J^\prime
    \sum_l
    n_{l,\uparrow} n_{l,\downarrow}
    \end{equation}

Analogously to the derivation leading to Eq.~\eqref{eq_1Ddensitydensitystaggered}, it is straightforward to show that a checkerboard modulation of the transverse field, $-\sum_{\mathbf{r}} g_\mathbf{r}\sigma^x_{\mathbf{r}}$ with $g_\mathbf{r}= g( 1+(-)^{r_x+r_y}\varepsilon)$, gives rise to a staggered fermion hopping,
\begin{equation}
\label{eq_1Dnnhoppingstaggered}
    H_{\rm 1D} =
    - g\sum_{j=1}^{L-1} \left( 1+(-)^{j}\varepsilon \right) \left(c^\dagger_j c_{j+1} + {\rm H.c.}\right) \, ,
\end{equation}
i.e., a Su-Schrieffer-Heeger chain.
Combining Eqs.~\eqref{eq_1Dnnhoppingstaggered} and~\eqref{eq_1Ddensitydensitystaggered} and choosing $\varepsilon=\delta$ we obtain a dimerized interacting-fermion Hamiltonian, equivalent to a bond-dimerized XXZ spin chain
\begin{equation}
\label{eq_1Dbonddimerized}
       H_{\rm bdXXZ} =
     \sum_{j=1}^{L-1} \left( 1+(-)^{j}\varepsilon \right) \, h^{\text{XXZ}}_{j,j+1}
    ,
    \end{equation}
where we used the shorthand $h^{\text{XXZ}}_{j,j+1} \equiv  -\frac g2 \left(\tau^x_j \tau^x_{j+1} + \tau^y_j \tau^y_{j+1} \right)
    +J^\prime \tau^z_j \tau^z_{j+1} $ [cf. Eq.~\eqref{eq_1DXXZ}].
Lastly, we observe that a checkerboard pattern in the nearest-neighbor Ising coupling, $- J \sum_{\mathbf{r}}  \left(\sigma^z_{\mathbf{r}} \sigma^z_{\mathbf{r}+\mathbf{e}_x}+\left(1+(-)^{r_x+r_y} h\right) \,\sigma^z_{\mathbf{r}} \sigma^z_{\mathbf{r}+\mathbf{e}_y} \right)$  gives rise to a staggered {chemical} potential for the fermions (or staggered magnetic field for the spins),
\begin{equation}
\label{eq_1Dpotentialstaggered}
    H_{\rm 1D} =
    -2Jh \sum_{j=1}^{L-1} (-)^{j} c^\dagger_j c_{j}  \, .
\end{equation}

\begin{table*}
\centering
\begin{tabular}{cc}
\toprule
2D quantum magnet \hspace{0.25cm} & Interface Hamiltonian \\
\midrule
$H_{\rm{2D}}= - J \sum_{\langle \mathbf{r},\mathbf{r'}\rangle} \sigma^z_\mathbf{r} \sigma^z_\mathbf{r'}$ \hspace{0.25cm} & ---
\\
\midrule
$- g \sum_\mathbf{r}  \sigma^x_\mathbf{r}$
\hspace{0.25cm}&
$-  \frac g2 \sum_{j} \Big(\tau^x_{j} \tau^x_{j+1} + \tau^y_{j} \tau^y_{j+1} \Big)$
\\
$-g^\prime \sum_{\mathbf{r}}\left(\sigma^x_{\mathbf{r}}\sigma^x_{\mathbf{r}+\mathbf{e}_x}
+
\sigma^x_{\mathbf{r}}\sigma^x_{\mathbf{r}+\mathbf{e}_y}\right)$
\hspace{0.25cm}&
$-  \frac {g^\prime}2 \sum_{j} \Big(\tau^x_{j} \tau^x_{j+2} + \tau^y_{j} \tau^y_{j+2} \Big)$
\\
$-g \varepsilon \sum_{\mathbf{r}} (-)^{r_x+r_y}  \sigma^x_{\mathbf{r}} $
\hspace{0.25cm}&
$ - \frac{g }2 \varepsilon \sum_{j} (-)^{j}  \Big(\tau^x_j \tau^x_{j+1} + \tau^y_j \tau^y_{j+1}\Big)$
\\
$-J^\prime \,\sum_{\mathbf{r}}\left(\sigma^z_{\mathbf{r}}\sigma^z_{\mathbf{r}+\mathbf{e}_x+\mathbf{e}_y}+\sigma^z_{\mathbf{r}}\sigma^z_{\mathbf{r}+\mathbf{e}_x-\mathbf{e}_y}\right)$
\hspace{0.25cm}& $+ J^\prime \,  \sum_{j} \tau^z_j {\tau^z_{j+1}}$
 \\
$ - J^\prime \delta \,\sum_{\mathbf{r}}  (-)^{r_x+r_y}\left( \sigma^z_{\mathbf{r}}\sigma^z_{\mathbf{r}+\mathbf{e}_x+\mathbf{e}_y}+\sigma^z_{\mathbf{r}}\sigma^z_{\mathbf{r}+\mathbf{e}_x-\mathbf{e}_y}\right) $
\hspace{0.25cm}& $ +J^\prime \delta \sum_{j} (-)^{j} \tau^z_j \tau^z_{j+1} $ \\
$ - Jh \sum_{\mathbf{r}}  (-)^{r_x+r_y} \,\sigma^z_{\mathbf{r}} \sigma^z_{\mathbf{r}+\mathbf{e}_y} $
\hspace{0.25cm}& $ -J h \sum_{j} (-)^{j} \tau^z_{j} $
\\
\bottomrule
\end{tabular}
\caption{
Correspondence between  bulk quantum perturbations to the 2D classical Ising magnet and the resulting 1D quantum spin-chain interactions describing interface dynamics to lowest order.
}
\label{tab_summaryIsing}
\end{table*}

In Table~\ref{tab_summaryIsing} we summarize the dictionary between 2D bulk perturbations and the corresponding effective 1D Hamiltonians governing interface quantum fluctuations at the leading order in perturbation theory.
For finite perturbations,  $H_{\rm 1D}$ features additional corrections
of order $\mathcal{O}(\lambda^n/J^{n-1})$ with
range proportional to $n$, for $n=2,3,\dots$.
Their effect can be straightforwardly included in both analytical and numerical calculations.
In Secs.~\ref{sec_phases} and~\ref{sec_neq} we will analyze the quantum phase structure and non-equilibrium dynamics of $H_{\rm 1D}$, and we will discuss the robustness of the resulting static and dynamical universal behaviors to all orders, i.e.,
for \textit{finite} perturbations to Eq.~\eqref{eq_classicalIsing} as far as an effective interface Hamiltonian $H_{\rm 1D}$ can be microscopically derived through the Schrieffer-Wolff transformation.

\subsection{Unveiling the 1D description}

\label{sec_unveiling}

The nature of low-energy quantum fluctuations and dynamics at the interface is revealed by \textit{imaging} the 2D spin lattice, i.e., simultaneously measuring all spin operators $\sigma^z_{\mathbf{r}}$. Each measurement collapses the system onto a classical spin configuration.

For \textit{infinitesimal} quantum perturbations, i.e., $J/\lambda\to\infty$, the 2D wave function lives in the exact ``holographic'' subspace mapped to 1D Fock space, as illustrated in Fig.~\ref{fig_mapping}(b,c).
Each classical configuration therein is characterized by the height field~$\varphi_j$.
The \textit{statistics} of 2D wave function snapshots thus directly reveal charge density fluctuations in the corresponding 1D wave function. In Sec.~\ref{sec_phases} and~\ref{sec_neq} we will show that these statistics fully characterize quantum phase structure and universal curvature-driven quantum dynamics, respectively.

For \textit{finite} quantum perturbations $\lambda \ll J < \infty$, however,  the exact ``holographic'' wave function described by the underlying 1D Hamiltonian emerges upon undoing the Schrieffer-Wolff transformation $U_{\rm SW}=e^{i \frac \lambda J S^{(1)} }e^{i \frac {\lambda^2} {J^2} S^{(2)} } \cdots$, as illustrated in Fig.~\ref{fig_mapping}(a,b). The latter may be viewed as a sequence of increasingly-short-time evolutions with increasingly-long-range hermitian generators $S^{(n)}$, or, effectively, a finite-depth quantum circuit. This dressing produces short-range-correlated local defects on top of the ``holographic'' wave function, with amplitudes of order $\mathcal{O}\left(\lambda/J\right)$.
As an experiment images the 2D spin lattice in the computational basis, the ``holographic'' configurations appear along with bulk and interface defects, with surface and linear density, respectively, of order $\mathcal{O}\left((\lambda/J)^2\right)$.
While the ``holographic'' description of static and dynamical behavior of the interface is well-defined for finite perturbations, it is natural to ask to which extent its dressing by $U_{\rm SW}$ may obscure experimental access.
Through numerical simulations of fairly large 2D spin lattices, in Sec.~\ref{sec_qusim} we will provide a practical demonstration that the dressing can be circumvented in experiment, and interface observations can be fully matched with ``holographic'' predictions.

The mismatch between computational and Schrieffer-Wolff-transformed basis may alternatively be removed through a ramp protocol in the perturbation strength, i.e., dynamically increasing $J(t) \to \infty$ while keeping $\vec\lambda$ fixed before imaging the system. Along such a ramp, the instantaneous effective Hamiltonian $H_{\rm 1D}(t)$ approaches the well-defined finite limit $\sum_\alpha  \lambda_\alpha P_{\rm 1D} V_{{\rm 2D},\alpha} P_{\rm 1D} $ as the higher-order corrections to $H_{\rm 1D}$ are progressively turned off [cf. Eq.~\eqref{eq_HSW}], and the Schrieffer-Wolff-transformed basis reduces to the computational basis $U_{\rm SW}(t)\to\mathbb{1}$.
Our goal is to perform the latter rotation while preserving the encoded 1D wave function we wish to unveil. 
For infinitely slow ramps, however, the 1D wave function will follow the adiabatic ground state of $H_{1D}(t)$. If, however, the ramp speed is fast compared to the typical interface coupling $\lambda$ but slow compared to the bulk gap $J$, then the 1D wave function will be effectively frozen during the ramp while the subspace it lives in will adiabatically approach the ideal ``holographic'' subspace.

We note that both the approaches above rely on a separation of scales $\lambda\ll J$, and, thus, are not guaranteed to work for strong perturbation $\lambda\approx J$, even though the Schrieffer-Wolff transformation and hence the underlying ``holographic'' 1D description of the interface may still be well-defined. How to unveil in experiment the microscopic 1D description in this regime remains an interesting open question that we do not address further in this work.

\section{Interface quantum phase structure}
\label{sec_phases}

We will now explore the macroscopically distinct quantum states induced on an interface by variations in the microscopic details of the bulk 2D quantum-magnet Hamiltonian, through the lens of geometric fluctuations.

The height observable is related to the effective 1D description via the relations
\begin{equation}
\label{eq_height_lattice}
\varphi_j-\varphi_0
=
2\sum_{l=1}^{j}c^\dagger_lc_l-j
=
\sum_{l=1}^{j}\tau_l^z \, .
\end{equation}
Equation~\eqref{eq_height_lattice} suggests a close connection with field-theoretical bosonization~\cite{Giamarchi}, which we discuss in Sec.~\ref{sec:noncompact_height}.
Hence, in Secs.~\ref{sec_xxz}-\ref{sec_Symmetry-enriched}, we will analyze the interface height-field statistics by computing the ground state of representative 1D Hamiltonians numerically using the density matrix renormalization group (DMRG) algorithm~\cite{White_DMRG_PhysRevLett.69.2863}, and hence by perfect sampling of the optimized matrix-product state wave function in the $\tau^z$-basis~\cite{Fishman_2022}.
These numerical calculations, in full agreement with field-theoretical bosonization predictions, support a ``holographic'' characterization of interface quantum phases and phase transitions. While throughout these sections we work with the lowest-order truncation of the Schrieffer-Wolff transformation $H_{\rm SW} \approx \lambda H_{\rm 2D}^{(1)}$ and $U_{\rm SW} \approx \mathbb{1}$ --- hence tacitly assuming infinitesimal perturbations, i.e.,  $J/\lambda\to\infty$ --- in Sec.~\ref{sec_symmetries} we will crucially prove that the symmetries protecting the various phases are exact to all orders in $\lambda/J$, and, hence, persist to \textit{finite} perturbations. This result shows that our ``holographic'' understanding of the interface quantum phase structure applies to 2D quantum magnets with {finite} couplings.

For ease of presentation, we will start by considering interfaces oriented horizontally (as in Fig.~\ref{fig_mapping}), ``holographically'' described by spin chains with zero total magnetization. In Sec.~\ref{sec_tilt} we will generalize the discussion to arbitrary spatial orientations.

\subsection{Exact bosonization}
\label{sec:noncompact_height}

An interface configuration is specified by the integer-valued height field $\varphi_j$.
We will now argue that the representation of interface quantum dynamics in terms of $\varphi_j$ is related to the effective 1D fermionic or spin Hamiltonians derived in Sec.~\ref{sec_mapping} via an exact lattice version of field-theoretical \emph{bosonization}.
Within the latter framework~\cite{Giamarchi}, one starts from a 1D Hamiltonian and maps the system onto an effective quantum field theory which can be analyzed using renormalization group (RG) techniques, allowing one to determine the qualitative large-scale physics of the system.
 In the simplest case, this quantum field theory is expressed in terms of two canonically conjugate compact bosonic fields $\tilde\varphi(x)$ and $\tilde\vartheta(x)$ satisfying $[\tilde\varphi(y),\partial_x \tilde\vartheta(x)] =  (2i/\pi) \delta(x-y)$;
 $\tilde\vartheta(x)$ is related to the canonical momentum  $\pi_{\tilde\varphi}$  by $\pi_{\tilde\varphi} = (\pi/2) \partial_x \tilde \vartheta$.\footnote{Our choice of field normalization is motivated by the correspondence with interface configurations, see below.}
 The fields are related to the spin operators as
\begin{align}
		\tau^\pm_j &\approx \left.  e^{\pm i \pi \tilde\vartheta(x)}\left(  (-1)^j \A~  + \C \cos(\pi \tilde\varphi(x)) + \ldots \right) \right|_{x=j} \nonumber\\
		\tau^z_{ j} &\approx \left.  \partial_x \tilde\varphi(x) + (-1)^j \mathcal{B} \sin (\pi\tilde\varphi(x)) + \ldots \right|_{x=j}
        \label{eq:Bosonization_spin}
\end{align}
 where, consistent with standard convention~\cite{Giamarchi}, the fields' argument is promoted to a continuous position variable $x$ smoothly interpolating between the discrete lattice sites $j$, and $\A$, $\B$ and $\C$ are non-universal prefactors [recall that we are assuming zero magnetization here, corresponding to a horizontal  interface as in Fig.~\ref{fig_mapping}(a); the general case is discussed in Sec.~\ref{sec_tilt} below]. These equations allow us to express the microscopic 1D Hamiltonian in terms of smoothly varying fields (with the argument $x$) and rapidly oscillating components at the lattice scale (with the argument $j$).

The compact field $\tilde\varphi$, in particular, enters physical interactions and observables only through its spatial derivatives $\partial^m\tilde\varphi$ and harmonics $e^{i\pi\ell \tilde\varphi}$, with $\ell$, $m$ integers, such as
\begin{equation}
\tilde{\varphi}\equiv \tilde{\varphi}+2 \, .
\end{equation}
Comparing Eqs.~\eqref{eq_height_lattice} and~\eqref{eq:Bosonization_spin}
then allows us to identify
\begin{align}
     \varphi_j-\varphi_0=\sum_{l=1}^j \tau^z_l &\approx \int^{x=j}_0 dy~ \partial_y\tilde\varphi(y) \nonumber\\ &= \left.2w(0,x) + \tilde\varphi(x)\right|_{x=j} \, ,
     \label{eq_height}
\end{align}
where $w(x_1,x_2)\in\mathbb{Z}$ is defined as the winding number of the compact field $\tilde\varphi(x)$ in the interval $[x_1,x_2]$.
In particular, the total winding number across the chain is vanishing for zero total magnetization,  $\varphi_L-\varphi_0=0=2w(0,L)$.
The field on the right-hand side of Eq.~\eqref{eq_height} can be interpreted as a non-compact \emph{unwinding} of $\tilde\varphi(x)$,
\begin{equation}
    \varphi(x) - \varphi(0) \equiv 2w(0,x) +  \tilde{\varphi}(x) \, ,
\end{equation}
taking arbitrary values in $\mathbb{R}$. By construction, $w(0,x)$ records the accumulated charge to the left of $x$ and $\tilde\varphi$ describes compact charge fluctuations; we can always replace $\tilde\varphi(x)$ by $\varphi(x)$ in Eq.~\eqref{eq:Bosonization_spin} and hence in all expressions of 1D local physical observables.
Thus, \emph{$\varphi_j$ represents an exact lattice version of $\varphi(x)$}.
 We will continue to denote both the exact discrete lattice height and its smooth coarse-grained counterpart by $\varphi$, since the distinction will always be clear from whether the argument is the lattice coordinate $j$ or the continuum coordinate $x$.

In summary, the exact lattice height field $\varphi_j$ represents the physical position of the interface on the 2D lattice.
As such, it is clearly a \textit{non-compact} physical observable, since different integer values of $\varphi_j$ correspond to physically distinguishable interface configurations. On the other hand, the {dynamics} of the height field map microscopically onto those of a quantum spin chain, which, by standard field-theoretic bosonization, is naturally described in terms of a {\it compact} continuum field $\tilde{\varphi}(x)$. To make the connection between interface quantum dynamics and field-theoretical bosonization more intuitive,
we recall that the exact mapping between spin and height configurations only involves the local {\it slopes}, i.e., differences in the heights between adjacent sites. A real-valued representative of the compact boson is obtained by fixing both the overall additive constant $\varphi_0$ and the winding sector. By Eq.~\eqref{eq_height}, this representative is naturally identified with a coarse-grained version of the exact height field. Although two representatives differing by $\varphi_0$ correspond to the same configuration in the bosonized spin-chain description, they remain physically distinct interface configurations. However, there is no contradiction because the height field itself is not a local operator in the spin variables: it is reconstructed by integrating the local slopes (equivalently, the charge $\tau^z$). Thus, the compactification reflects the local operator content of the spin-chain description, whereas the non-compact physical height field survives as a nonlocal observable.
{In Appendix~\ref{app_soft} we provide a more detailed discussion of the subtleties of exact bosonization, including a discussion of how the renormalized  quantum field theory describing  macroscopic interface physics  in terms of the continuum field $\varphi(x)$ can be directly obtained  from the microscopic lattice Hamiltonian describing the exact height field $\varphi_j$ dynamics.}

From an experimental point of view, the natural non-compactness of the interface height field is a central advantage of ``holographic'' quantum simulation, as certain relevant non-local observables in~1D become directly measurable local observables in~2D. A primary example is given by measurements of $\varphi_j$ itself, which yield the charge full counting statistics of the encoded 1D system, opening the door to symmetry-resolved measurements.
Relatedly, Ising-symmetry breaking fields in 2D give rise to interactions involving a non-compact gauge field in 1D, opening the door to analog simulation of \textit{bona fide} continuum gauge field theories~\cite{lerose2024simulating}. We will return to experimental consequences in Sec.~\ref{sec_qusim}.

\subsection{Interface stiffness and roughness}
\label{sec_xxz}
\begin{figure*}
   \includegraphics[width=\linewidth]{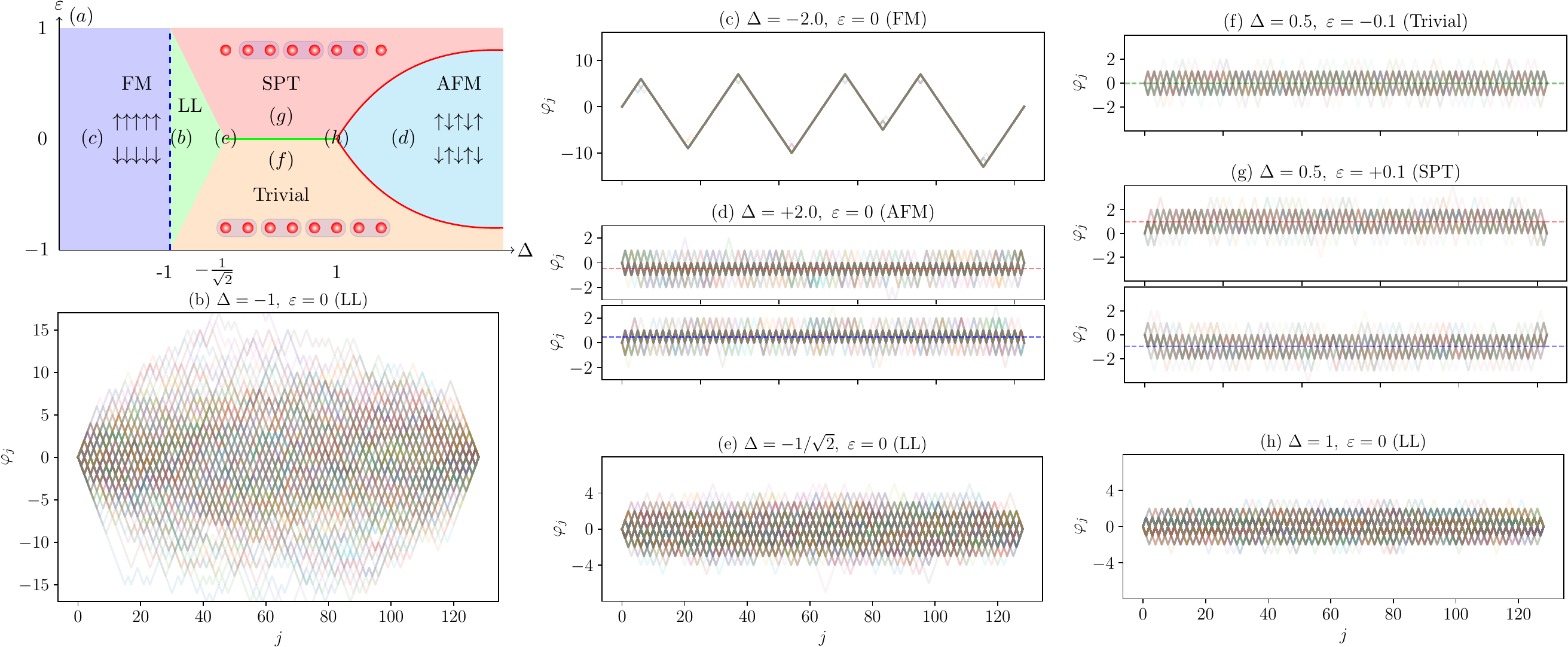}
   \caption{
   \textbf{Interface quantum phase structure (I).}
   Panel (a): Schematic phase diagram of the effective 1D Hamiltonian~\eqref{eq_1Dbonddimerized}. Panels (b-h): Interface fluctuation  statistics computed from the 1D ground-state wave function in various phases: (b,e,h) Luttinger-liquid (LL) phase for three values of the anisotropy parameter $\Delta=2J^\prime/g$; (c) Ferromagnetic and (d) antiferromagnetic (N\'{e}el) phases; (f)  trivial and (g) Haldane SPT gapped phases. Degenerate ground states are shown for the N\'{e}el and Haldane SPT  phases. Horizontal dashed lines highlight field-theoretical bosonization predictions. Here $L=128$ and the plots are made by overlapping 50 (200) wave-function snapshots for gapped (gapless) states. \label{fig:combined_dimer}}
\end{figure*}

We will now study the interface phase structure by analyzing height fluctuations as a function of the various 2D bulk perturbations. For this, we will use field-theoretical bosonization and DMRG for the corresponding 1D Hamiltonian.

As the height field plays the role of counting field for 1D charge statistics, we anticipate a simple and sharp one-to-one correspondence between quantum roughening transitions for 2D interfaces and underlying conductor-insulator transitions in $U(1)$-symmetric 1D systems. Indeed, in short-range correlated gapped phases, fluctuations of the total charge contained in a large region  are confined to finite distances of the order of the correlation length from the boundary, leading to finite interface height fluctuations. On the other hand, in long-range correlated gapless phases described by (1+1)D conformal field theory, scale invariance implies that charge fluctuations diverge with the size of the region, leading to interface roughness.

We will illustrate this correspondence starting from the archetypal XXZ spin chain in
Eq.~\eqref{eq_1DXXZ}, generated by a weak transverse field $g>0$ and weak beyond-nearest-neighbor couplings $J^\prime$ in the 2D Ising magnet. The result of field-theoretical bosonization of this spin chain reads
\begin{multline}
    H_{QFT} \approx \frac{\pi v}{2} \int dx \left( \frac{(\partial_x \varphi)^2}{4K} + K (\partial_x \vartheta)^2 \right) \\+ \frac{J' \mathcal{B}^2}{g} \int dx~ \cos(2\pi\varphi). \label{eq:Bosonization_XXZ}
\end{multline}
Here $v$ is the Luttinger velocity, which sets the overall energy scale, and $K$ is the Luttinger parameter, which controls the scaling dimensions $X_{mn}$ of the operators~$\mathscr{O}_{mn}= e^{i\pi(m\varphi + n\vartheta)}$,
\begin{equation}
X_{mn}  = m^2 K + \frac{n^2}{4K} \, ,
\end{equation}
through which we can analyze various perturbations.  For the XXZ chain, the values of $K$ and $v$ can be explicitly calculated as a function of the standard anisotropy parameter $\Delta=2J^\prime/g$ through Bethe ansatz~\cite{Haldane_BetheLL_1981153}:
\begin{equation}
		K = \frac{\pi}{2  \arccos (-\Delta)} , ~ v =  \frac{2gK}{(2K-1)} \sin \left(\frac{\pi}{2K}\right) \, . \label{eq:Bethe_Luttinger}
\end{equation}
For $\Delta <1$ the leading perturbation $\cos(2\pi\varphi)$ to the quadratic action is irrelevant (in the RG sense) and the spin chain belongs to the gapless Luttinger liquid (LL) phase~\cite{Haldane_1981_LuttingerLiquid}, while for $\Delta>1$ it is relevant and the action flows to a gapped fixed point under RG, corresponding to the Néel phase. The transition occurs
at the isotropic ($SU(2)$-symmetric) Heisenberg point of the XXZ spin chain.

In \cref{fig:combined_dimer}(b,e,g) we show how the Luttinger liquid phase appears in terms of height field $\varphi_j$ fluctuations for various values of $\Delta$.
\begin{figure*}
    \centering
    \includegraphics[width=\linewidth]{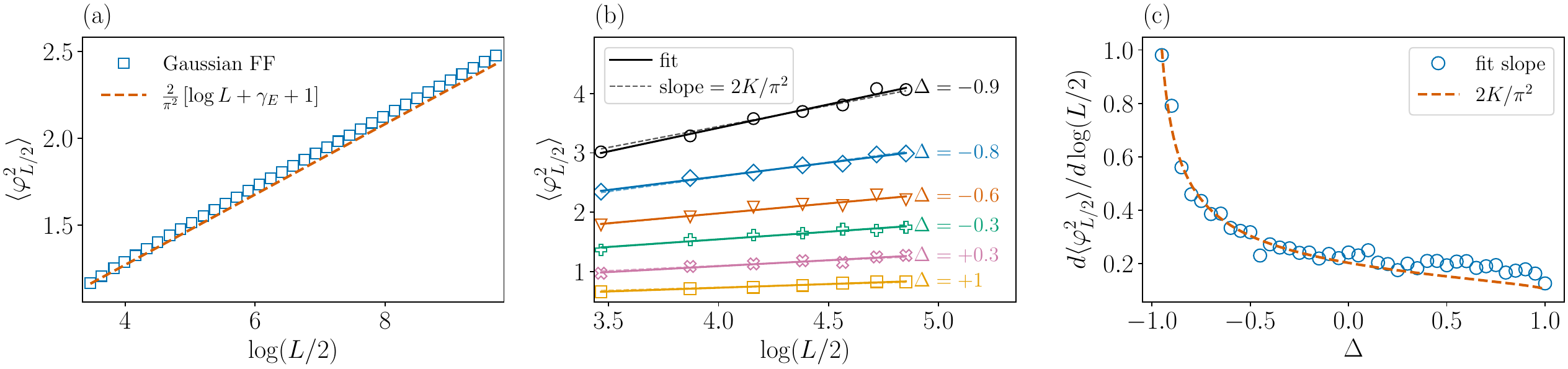}
    \caption{   \textbf{Roughness scaling law \eqref{eq:LL_fluctuation_bosonization}.} Panels (a,b): Mid-point variance of the interface height $\langle \varphi^2_{L/2}\rangle$ vs interface length, numerically computed from the effective 1D XXZ spin chain ground state, for (a) the free-fermion point $J^\prime = 0$  and (b)  general values of the effective anisotropy $\Delta=2J^\prime/g$. Panel (c): illustration of the scaling law in \cref{eq:LL_fluctuation_bosonization}. For the free-fermion point, we overlay the numerically exact result of $\langle \varphi^2_{L/2}\rangle$ from Gaussian techniques with the known large-size asymptotics $\langle \varphi_{L/2}^2\rangle = \frac{2}{\pi^2}\left[ \log (L/2) + \log2 +\gamma_E + 1 + o(1) \right] $,
where $\gamma_E$ is the Euler-Mascheroni constant~\cite{Eisler_2006}. For the general case (b,c), $\langle \varphi^2_{L/2}\rangle$ is computed by averaging over $10^5$ ground state wave-function snapshots from DMRG.
}
    \label{fig:varVersusKN}
\end{figure*}
 This is a rough phase of the interface, as the amplitude of  fluctuations diverges with system size, with a scaling law that depends parametrically on $\Delta$  (through the Luttinger parameter).  Using the bosonization framework, it can be shown that~\cite{Giamarchi}
\begin{equation}
    \langle \varphi(x) \rangle = 0,  ~  \langle \varphi^2({L/2}) \rangle  \sim \frac {2K} {\pi^2} \log(L/2) \label{eq:LL_fluctuation_bosonization}.
 \end{equation}
A numerical calculation confirming the scaling law in \cref{eq:LL_fluctuation_bosonization} is shown in \cref{fig:varVersusKN}, where we further support this conclusion through an exact calculation at the free-fermion point $\Delta=0$. This scaling behavior of interface fluctuation statistics signals the imprint of the underlying Luttinger-liquid nature of this phase.

For $|\Delta|>1$, the XXZ spin chain  transitions to spontaneous-symmetry-breaking gapped phases. For $\Delta<-1$, the spin chain belongs to a ferromagnetic (FM) phase which is separated from the Luttinger liquid phase by a first-order phase transition. In this phase, energy is locally minimized by the two fully magnetized product states with $\tau^z_j=+1$ or $\tau^z_j=-1$ for all $j$, respectively. When we are restricted to the zero magnetization sector,
ferromagnetism manifests itself in the formation of large magnetized domains, i.e., phase separation. In terms of interface configurations, this phenomenon appears in the form of straight macroscopic segments randomly selected by DMRG runs, as shown in \cref{fig:combined_dimer}(c).

For $\Delta>1$ on the other hand, the spin chain belongs to a N\'{e}el antiferromagnetic (AFM) phase, with two degenerate ground states distinguished by the sign of the staggered magnetization $\sum_j (-)^j \tau^z_j$. In terms of interface behavior, this phase manifests itself as pinning to the lattice, with weak fluctuations around the horizontal zigzag or anti-zigzag classical configurations, as shown in \cref{fig:combined_dimer}(d). This is a  `stiff' phase of the interface, as the amplitude of fluctuations stays finite as $L\to\infty$.

Remarkably, field-theoretical bosonization predicts the correct behavior at the lattice scale:
When $\cos(2\pi\varphi)$ becomes relevant in \cref{eq:Bosonization_XXZ}, it is minimized by $\langle\varphi \rangle = \pm \frac{1}{2}, \pm \frac{3}{2}, \ldots$. Due to the pinned boundary conditions, the two values $\langle\varphi \rangle  = \pm \frac{1}{2} $ are selected. This agrees with the mean value of the exact lattice height field computed numerically, as shown by the dashed lines in \cref{fig:combined_dimer}(g). The finite correlation length results in a finite variance $\langle \varphi_j^2 \rangle - \langle \varphi_j \rangle^2$ of interface fluctuations in the two symmetry-broken states, which decays to zero for large $\Delta$.

\subsection{ Trivial and topological stiff phases}

In order to access a richer interface phase diagram, we will now consider the staggered 2D bulk interactions leading to the bond-dimerized XXZ spin chain in Eq.~\eqref{eq_1Dbonddimerized}.
This Hamiltonian can be bosonized as follows
\begin{multline}
    H_{QFT} \approx \frac{v\pi}{2} \int dx \left( \frac{(\partial_x \varphi)^2}{4K} + K (\partial_x \vartheta)^2 \right) \\+  \varepsilon \A \C  \int dx  \cos (\pi\varphi)+ \frac{J' }{g} \mathcal{B}^2 \int dx~ \cos(2\pi\varphi) \label{eq:Bosonization_XXZ_dimerization}
\end{multline}
Dimerization introduces a perturbation proportional to $\cos (\pi\varphi)$, whose scaling dimension is $K$. This operator is relevant for  $K<2$, i.e., for $\Delta > -1/\sqrt{2}$. In this regime the Luttinger liquid of the XXZ chain is unstable to this perturbation, and the RG flow  leads to a gapped symmetry preserving phase.
The phase diagram of Eq.~\eqref{eq_1Dbonddimerized} is shown in \cref{fig:combined_dimer}(a). Depending on the sign of $\varepsilon$, the gapped phase is either trivial ($\varepsilon<0$) or a symmetry protected topological (SPT) phase ($\varepsilon>0$). For  $|\varepsilon|=1$, these ground states take the form of a product of spin-singlet dimers occupying the odd or even bonds of the chain, respectively, as schematically shown in \cref{fig:combined_dimer}(a).

We can use bosonization to study how these phases are manifested in interface physics. When the perturbation $\cos(\pi\varphi)$ is relevant in \cref{eq:Bosonization_XXZ_dimerization}, bulk minima occur at $\langle \varphi \rangle$ being pinned to \emph{all} even (odd) integers for $\varepsilon<0~(>0)$, since $\varphi$ is not compact. However, the boundary condition $\varphi_0=\varphi_L=0$ energetically picks a unique vacuum with $\langle \varphi \rangle = 0$ for the trivial phase  when $\varepsilon<0$. For $\varepsilon>0$ on the other hand, there are two minimal energy non-compact lifts, $\langle \varphi \rangle = \pm 1$.  The pinned values of $\varphi$ for the trivial and SPT phases are verified numerically in \cref{fig:combined_dimer}(e,f).  The interface configurations for the SPT phase contains opposite half-kinks at the two boundaries. Since a change $\Delta\varphi$ carries spin
\begin{equation}
S^z_{\rm edge}=\frac{\Delta\varphi}{2},
\end{equation}
the two lifts correspond to the edge configurations
\begin{equation}
\left(S^z_L,S^z_R\right)
=
\left(+\frac12,-\frac12\right),
\qquad
\left(-\frac12,+\frac12\right).
\end{equation}
They are precisely the two degenerate ground states in the zero-magnetization sector.\footnote{The total degeneracy of the spin chain ground state is four, due to the two-fold degeneracy on each end. However, the restriction to the zero magnetization sector selects only two of the four states as relevant interface states. \label{footnote_degeneracy}}

Higher odd lifts, such as $\langle \varphi \rangle =\pm3$, contain an additional full kink carrying integer spin and are therefore excited states, not additional ground states. For $\varepsilon<0$, the minimum $\langle \varphi \rangle =0$ matches the pinned boundary without any kink and gives a unique trivial-dimer ground state. Configurations centered around $\varphi=\pm2,\pm4,\ldots$ require additional integer-spin kinks at the boundaries and are again excited states.

\subsection{Valence-bond-solid
stiff phase}

A fertile setting to study an even richer phase structure is by considering spin ladder models or, equivalently, next-nearest neighbor interactions.
In Sec.~\ref{sec_mapping} we identified bulk 2D perturbations that generate 1D interactions
of the form
\begin{multline}
   \label{eq_ladder}
   H_{\text{ladder}} = H_{\text{bdXXZ}} + \frac{g'}2\sum_{j} (\tau^x_{j} \tau^x_{j+2} + \tau^y_j \tau^y_{j+2}
   )  \, ,
\end{multline}
where $H_{\text{bdXXZ}}$ is given by \cref{eq_1Dbonddimerized} and we assume $g^{\prime}>0$ for definiteness.

For sufficiently small $J'$,
as we increase the ratio $\alpha = g'/g$ the original Luttinger liquid phase of the XXZ spin chain transitions to a valence-bond-solid (VBS) phase. On the spin chain, this phase is characterized by a bulk ground-state degeneracy corresponding to spin singlets occupying even and odd bonds, respectively. In other words, the trivial and SPT ground states discussed previously appear as part of the same ground-state manifold.
On the bosonized Hamiltonian, the next-nearest neighbour coupling modifies \cref{eq:Bosonization_XXZ} as
\begin{multline}
    H_{QFT} \approx \frac{\pi v}{2} \int dx \left( \frac{(\partial_x \varphi)^2}{4K} + K (\partial_x \vartheta)^2 \right) \\+ \frac{(J'-g') \mathcal{B}^2}{g} \int dx~ \cos(2\pi\varphi). \label{eq:Bosonization_XXZ_nnn}
\end{multline}
The VBS phase corresponds to $g'>J'$ and $\cos(2\pi \varphi)$ being relevant. If $\varphi$ were compact, this picks out $\langle \varphi \rangle = 0,1$ as the degenerate minima corresponding to the trivial and topological dimer states.

For the physical interface ground state, however, the term $-\cos(2\pi\varphi)$ is minimized by $\langle{\varphi}\rangle$ being pinned to any integer value. However, the fixed boundary condition  picks out a unique minimum $\langle{\varphi}\rangle = 0$. In the spin-chain picture, this lifted degeneracy corresponds to the boundary conditions explicitly breaking translation invariance, which favors one of the two dimer configurations.

The boundary-condition-selected VBS state and the trivial dimer generated by $-\cos(\pi\varphi)$ therefore have the same large-scale interface ground-state profile: both are flat and pinned at $\langle\varphi\rangle=0$. This is the interface realization of the correspondence identified by den Nijs and Rommelse: in the statistical mechanics of crystal surfaces, the VBS phase is a ``disordered flat'' phase, whose height fluctuations remain bounded despite a disordered array of microscopic steps~\cite{dennijs1989preroughening,santoro1994disordered}. Thus, macroscopic flatness does not by itself imply a trivial phase. Here, the VBS and trivial dimer are distinguished by their translation properties and by the structure and symmetry quantum numbers of their lowest-lying excitations; with periodic boundary conditions, the distinction is also manifest in the spontaneous translation-breaking degeneracy of the VBS. The role of compactification, pinned boundary conditions, and the Lieb-Schultz-Mattis (LSM) constraint in this ground-state counting is discussed in detail in Appendix~\ref{app_soft}.

\subsection{
Symmetry-enriched rough phases}

\label{sec_Symmetry-enriched}

\begin{figure}
    \centering
    \includegraphics[width=.8\linewidth]{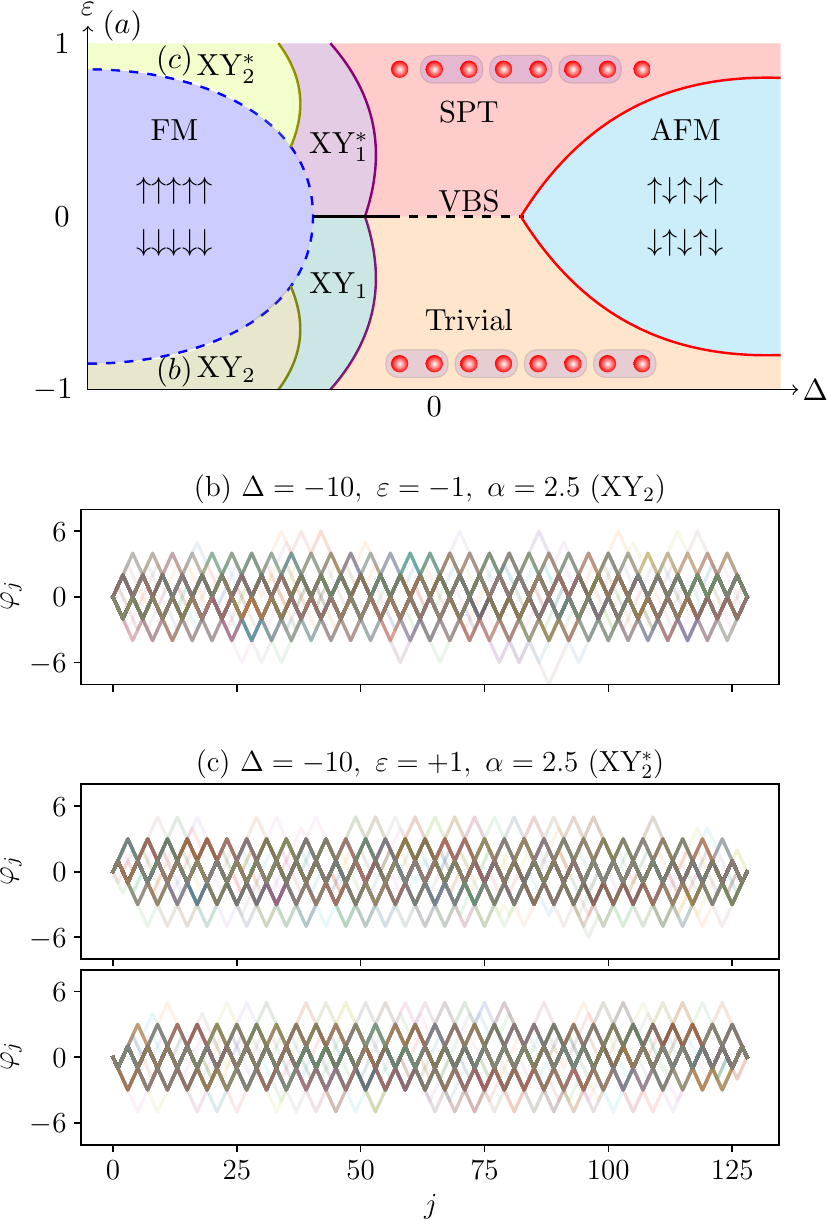}
    \caption{\textbf{Interface quantum phase structure (II).} Panel (a): Schematic phase diagram of the effective 1D Hamiltonian~\eqref{eq_ladder} for $g'/g = \alpha \sim 2.5$.
   Panels (b,c): Interface fluctuation statistics  computed from the 1D ground-state wave function in the rough trival (XY$_2$) (b) and non-trivial (XY$_2^*$) (c) symmetry-enriched critical phases. XY$_2^*$ has degenerate ground states with distinct edge modes that are shown. Here $L=128$ and the plots are made by overlapping 200 wave-function snapshots.
    }
    \label{fig:XXZ_ladder}
\end{figure}

Several other interesting states appear for large next-nearest-neighbour coupling $\alpha=g'/g$~\cite{AP_TLL_PhysRevB.108.245135} including distinct `symmetry enriched' Luttinger liquids. These phases can be studied using a more complex two-component bosonization framework~\cite{Schulz_Spin1_PhysRevB.34.6372,AP_TLL_PhysRevB.108.245135,AP_UCMulti_PhysRevLett.130.256401,AP_UC_Pump} which we will not review here. As a demonstration, let us consider a Luttinger liquid phase with underlying coexisting ferromagnetic correlations, dubbed XY$_2$~\cite{Schulz_Spin1_PhysRevB.34.6372,AP_UCMulti_PhysRevLett.130.256401,AP_UC_Pump}. Numerical simulation of this phase in the interface variables is shown in \cref{fig:XXZ_ladder}(b). The underlying ferromagnetic correlations which occurs in the spin chain through alignment of neighboring spins is visible in the interface wave function as `doubling' of the fluctuating units compared to the ordinary rough phase in \cref{fig:combined_dimer}(b,e,g). Furthermore, recent work has also shown the existence of a `topological' counterpart to XY$_2$, labeled XY$_2^*$ which has protected edge modes~\cite{AP_TLL_PhysRevB.108.245135}. \cref{fig:XXZ_ladder}(c) shows the corresponding interface phase. In spite of divergent fluctuations in the bulk associated with roughness, the topological degeneracy results in a net transversal displacement of the interface, which is more evident from the behavior near the boundaries.

\subsection{Exact symmetries and phase robustness}

\label{sec_symmetries}

\begin{table*}[t]
	\centering
	\small
	\setlength{\tabcolsep}{3pt}
	\renewcommand{\arraystretch}{1.12}
	\begin{tabular}{lll}
		\toprule
		\parbox[t]{0.18\textwidth}{\raggedright Symmetry}
		&
		\parbox[t]{0.21\textwidth}{\raggedright Action on the 1D spins}
		&
		\parbox[t]{0.57\textwidth}{\raggedright Action on the 2D spins and microscopic origin} \\
		\midrule
		\parbox[t]{0.18\textwidth}{\raggedright $U(1)$ spin rotation}
		&
		\parbox[t]{0.21\textwidth}{\raggedright
			$\tau^\pm_j\mapsto e^{\pm i\chi}\tau^\pm_j$,
			$\tau^z_j\mapsto\tau^z_j$}
		&
		\parbox[t]{0.57\textwidth}{\raggedright
			The boundary conditions impose the conservation of the total magnetization as a hard geometric constraint.  There is no simple onsite action on the
			microscopic spins $\vec{\sigma}_{\mathbf r}$.
			            } \\
		\addlinespace
		\parbox[t]{0.18\textwidth}{\raggedright $\mathbb Z_2^R$ spin reflection}
		&
		\parbox[t]{0.21\textwidth}{\raggedright
			$\tau^\pm_j\mapsto\tau^\mp_j$,
			$\tau^z_j\mapsto-\tau^z_j$}
		&
		\parbox[t]{0.57\textwidth}{\raggedright
			$\sigma^x_{\mathbf r}\mapsto\sigma^x_{\mathbf r'}$ and
			$\sigma^{y,z}_{\mathbf r}\mapsto-\sigma^{y,z}_{\mathbf r'}$, where
			$\mathbf r'$ is reflected across the interface baseline.  This is the
			microscopic Ising spin flip combined with lattice reflection.  } \\
		\addlinespace
		\parbox[t]{0.18\textwidth}{\raggedright $\mathbb Z_2^P$ lattice parity}
		&
		\parbox[t]{0.21\textwidth}{\raggedright
			$\vec{\tau}_j\mapsto
			\vec{\tau}_{-j+1}$}
		&
		\parbox[t]{0.57\textwidth}{\raggedright
			$\sigma^x_{\mathbf r}\mapsto\sigma^x_{\mathbf r'}$ and
			$\sigma^{y,z}_{\mathbf r}\mapsto-\sigma^{y,z}_{\mathbf r'}$, where
			$\mathbf r'$ is obtained by a twofold lattice rotation about the
			appropriate bond center.  } \\
		\addlinespace
		\parbox[t]{0.18\textwidth}{\raggedright
			One-site translation}
		&
		\parbox[t]{0.21\textwidth}{\raggedright
			$\vec{\tau}_j\mapsto
			\vec{\tau}_{j+1}$}
		&
		\parbox[t]{0.57\textwidth}{\raggedright
			Primitive translations of the square lattice, $\vec{\sigma}_{\mathbf r}\mapsto
			\vec{\sigma}_{\mathbf r+\mathbf e_x}$ or
			$\vec{\sigma}_{\mathbf r+\mathbf e_y}$.  } \\
		\addlinespace
		\parbox[t]{0.18\textwidth}{\raggedright
			Two-site translation}
		&
		\parbox[t]{0.21\textwidth}{\raggedright
			$\vec{\tau}_j\mapsto
			\vec{\tau}_{j+2}$}
		&
		\parbox[t]{0.57\textwidth}{\raggedright
			Diagonal translation of the square lattice, $\vec{\sigma}_{\mathbf r}\mapsto
			\vec{\sigma}_{\mathbf r+\mathbf e_x+\mathbf e_y}$.  } \\
		\bottomrule
	\end{tabular}
    \caption{\label{tab:interface_symmetries}
		Symmetries of the effective 1D Hamiltonian and their origin
		in the parent 2D spin lattice.}
\end{table*}

\begin{table}[t]
	\centering
	\small
	\setlength{\tabcolsep}{3pt}
	\renewcommand{\arraystretch}{1.4}
	\begin{tabular}{ll}
		\toprule
		\parbox[t]{0.36\columnwidth}{\raggedright Phase}
		&
		\parbox[t]{0.51\columnwidth}{\raggedright Sufficient symmetry set} \\
		\midrule
		\parbox[t]{0.36\columnwidth}{\raggedright LL}
		& \parbox[t]{0.51\columnwidth}{\raggedright $U(1)$} \\
		\parbox[t]{0.36\columnwidth}{\raggedright FM}
		& \parbox[t]{0.51\columnwidth}{\raggedright $\mathbb Z_2^R$} \\
		\parbox[t]{0.36\columnwidth}{\raggedright AFM}
		& \parbox[t]{0.51\columnwidth}{\raggedright
			$\mathbb Z_2^R$ and either $\mathbb Z_2^P$ or one-site translation} \\
		\parbox[t]{0.36\columnwidth}{\raggedright Trivial and SPT}
		& \parbox[t]{0.51\columnwidth}{\raggedright
			$U(1)$ and $\mathbb Z_2^R$} \\
		\parbox[t]{0.36\columnwidth}{\raggedright VBS}
		& \parbox[t]{0.51\columnwidth}{\raggedright
			$U(1)$, $\mathbb Z_2^R$, and one-site translation} \\
		\parbox[t]{0.36\columnwidth}{\raggedright XY$_2$ and XY$_2^*$}
		& \parbox[t]{0.51\columnwidth}{\raggedright
			$U(1)$, $\mathbb Z_2^R$, $\mathbb Z_2^P$, and two-site translation} \\
		\bottomrule
	\end{tabular}
    \caption{\label{tab:phase_symmetries}
		Sufficient symmetry sets for the various interface phases in Figs.~\ref{fig:combined_dimer} and~\ref{fig:XXZ_ladder}.}
\end{table}

The interface phases discussed so far require particular symmetries to be present in the effective 1D
 Hamiltonian.
In \cref{tab:interface_symmetries} we collect a useful menu of such
symmetries,
and we trace their exact origin from the parent 2D spin system.
A sufficient set of microscopic symmetries for stabilizing each phase is reported in \cref{tab:phase_symmetries}.\footnote{These sets are not intended to be minimal: additional symmetries may be present, and in some cases alternative protecting subgroups are possible. In particular, the symmetry set $U(1)$, $\mathbb Z_2^R$, $\mathbb Z_2^P$, and two-site translation that
stabilizes and distinguishes XY$_2$ and XY$_2^*$~\cite{AP_TLL_PhysRevB.108.245135} is sufficient to stabilize
the entirety of the two phase diagrams in \cref{fig:combined_dimer}(a), \cref{fig:XXZ_ladder}(a) considered above.  One-site translation appears as an
enhanced symmetry on special submanifolds of the phase diagram, such as the dimerization-free line supporting the XXZ phase diagram and VBS phase.}

One may ask whether these symmetries of the effective 1D Hamiltonian
are merely accidental symmetries of the leading-order
Schrieffer-Wolff transformation, which get explicitly broken at higher orders.
In fact, we can formally prove that all of them are robust to all orders
in $H_{\rm 1D}$. Thus, they are  \textit{exact emergent symmetries} of
interface dynamics, revealed by undoing the Schrieffer-Wolff transformation $U_{\rm SW}$.

We have already discussed in Sec.~\ref{sec_H1D} how the $U(1)$ symmetry is an exact emergent symmetry.
The remaining symmetries can be inherited from exact spatial symmetries of the 2D Hamiltonian,
as detailed in \cref{tab:interface_symmetries}.
Consider one such exact symmetry $T$, which additionally preserves the
low-energy interface subspace of minimal-length domain-wall configurations.
By hypothesis,  $[T,V_{2D}]=0$ and $[T,H_{2D}^{(0)}]=0$ (hence, $[T,P_{\rm 1D}]=0$). Then, $T$ is preserved through the steps of the Schrieffer-Wolff block-diagonalization algorithm, as can already be seen in Eq.~\eqref{eq_HSW} for the lowest orders. It can be formally proven that $[T,H_{\rm 2D}^{(1)}]=0$ and $[T,S^{(1)}]=0$, as well as the induction step, i.e., that $[T,S^{(1,\cdots,n)}]=0$ implies $[T,S^{(n+1)}]=0$ and $[T,H_{\rm 2D}^{(n+1)}]=0$.
This shows that the inherited symmetry $T_{\rm 1D}=P_{\rm 1D} T P_{\rm 1D}$ constrains the effective
1D Hamiltonian of the interface at \textit{every} perturbative order, $[T_{\rm 1D},H_{\rm 1D}]=0$, thereby protecting the interface phase structure for finite perturbations. Moreover, since $[T,U_{\rm SW}]=0$ to all orders, the symmetry acts in the same way in the computational and Schrieffer-Wolff-transformed bases.

For the pinned boundary conditions used here, $\mathbb Z_2^R$ and $\mathbb Z_2^P$ are
exact when the two endpoints and the frozen boundary spins are chosen
symmetrically.  The pinned endpoints explicitly break translations only at
the boundaries; one-site or two-site translation remains an exact bulk
symmetry.\footnote{It  becomes a literal symmetry with periodic boundary conditions or
for infinite horizontal system size.}  Consequently, higher-order corrections
may only renormalize the phase boundaries and edge profiles, and symmetry-allowed finite-size edge
splittings may remain.

We can contrast the robustness of the symmetries discussed above to the emergent $SU(2)$ spin symmetry we have found for the special parameters values $J^\prime=\pm g/2$ in the effective lowest-order 1D Hamiltonian~\eqref{eq_1DXXZ}. In this case, there seems to be no exact bulk symmetry of the 2D quantum magnet from which the emergent 1D symmetry is inherited. Accordingly, in Sec.~\ref{sec_neq} we will show that this symmetry is accidental, as it gets explicitly broken at second order.

\subsection{General spatial orientations}
\label{sec_tilt}

Our analysis so far was restricted to horizontal interfaces, with endpoints pinned at equal height, as in Fig.~\ref{fig_mapping}(a).
For this case, we have shown that
infinitesimal bulk  perturbations with the simplest nontrivial periodicity --- a checkerboard strength pattern ---
may fully suppress roughness.
We will now extend our discussion to interfaces with a general orientation angle with respect to the lattice axes, and we will show that suitable spatially-modulated bulk perturbations can similarly suppress roughness.

Tilted interfaces are described by the 1D Hamiltonians derived in Sec.~\ref{sec_mapping} projected into the sector with filling factor $\rho=N/L$, or, equivalently, magnetization density $m=\sum_j\tau^z_j/L=2N/L-1$. The orientation angle $\Theta$ is related to the parameters $\rho$, $m$ by the equation
\begin{equation}
    \tan \Theta = m = 2\rho-1 \, .
\end{equation}

For simplicity, let us first consider the case of a transverse-field bulk perturbation, leading to a pure-hopping free-fermion Hamiltonian.
A homogeneous transverse field $g_\mathbf{r}=g$ yields a free-fermion dispersion relation $E(k)=-2g\cos k$ and hence a Fermi-sea ground state with Fermi momentum
\begin{equation}
\label{eq_kF}
    k_F=\pi\rho = \frac{\pi}{2}  (1+m) \, .
\end{equation}
This 1D state is gapless except for $\rho=0,1$, corresponding to a rough interface for all $\Theta\neq \pm \pi/4$. Interfaces with $\Theta= \pm \pi/4$ --- i.e. with orientation parallel to one of the lattice axes --- are instead stiff, corresponding to a completely empty or completely full energy band.
A horizontal $R$-periodic modulation in the transverse field  $g_\mathbf{r}=g_{\mathbf{r}+R\mathbf{e}_x}=g_{\mathbf{r}+\mathbf{e}_x-\mathbf{e}_y}$, induces an $R$-periodic modulation of the hopping $g_j=g_{j+R}$.
The bond-dimerization case $R=2$ [cf. Eq.~\eqref{eq_1Dnnhoppingstaggered}] opens a band gap at $k^*=\pm \pi/2$, i.e., at the Fermi momentum associated with  a horizontal interface. As we have already seen, this gap leads to interface stiffness for $\Theta=0$. For all other values $\Theta\neq 0,\pm\pi/4$ the interface is rough.
Generalizing the above argument, a spatial modulation of period $R$ opens band gaps at momenta $k^*=\pm\pi/R,\pm2\pi/R,\dots,\pm(R-1)\pi/R$. The corresponding interfaces with slopes
\begin{equation}
    \tan \Theta = 1, 1 - \frac2R, 1-\frac 4R, \dots,-1+\frac2R, -1 \label{eq:OYA_condition}
\end{equation}
become stiff, while for all other values of $\Theta$ they are rough.

The same scenario occurs for other kinds of horizontal $R$-periodic perturbations. For example, modulations of the Ising couplings $h_\mathbf{r}=h_{\mathbf{r}+R\mathbf{e}_x}=h_{\mathbf{r}+\mathbf{e}_x-\mathbf{e}_y}$ generalizing \cref{eq_1Dpotentialstaggered} lead to an $R$-periodic 1D single-particle potential $h_j=h_{j+R}$. This perturbation opens band gaps and thus makes the interface stiff for all the orientations given by \cref{eq:OYA_condition}.
Since $R$ is an arbitrary integer, we conclude that interfaces with an arbitrary \textit{rational} slope can become stiff (i.e., pinned to the lattice) under infinitesimal bulk perturbations with an appropriate commensurate periodic modulation.

To elucidate the effect of interactions we can once again use bosonization. For a general magnetization density $m$  spin operators are expressed in terms of  bosonic quantum fields as follows
	\begin{align}
	\tau^{\pm}_{ j} &\approx  (-1)^j e^{\pm i\pi \vartheta(x) }\left( \A  + \C \cos \left(\pi \varphi(x) + 2 k_F j\right) + \ldots \right),  \nonumber \\
	\tau^z_{ j} &\approx m +  \partial_x \varphi(x) +  \mathcal{B} \sin \left(\pi\varphi(x) + 2 k_F j\right) + \ldots  \, ,\label{eq:Bosonization_single_magfield}
	\end{align}
  				 with $k_F$ given by \cref{eq_kF}, thus generalizing \cref{eq:Bosonization_spin}.
     For definiteness, let us consider an $R$-periodic magnetic field $ H_{\rm 1D} = \sum_j h_j \tau^z_j$, with~$ h_j = h_{j+R}$. By Fourier transformation of  $h_j$,
     \begin{equation}
         h_j = \sum_{\ell = 0}^{R-1}\tilde{h}_\ell \cos\left(\frac{2 \pi \ell j}{R} + \chi_\ell \right),
     \end{equation}
     the most relevant terms this introduces in the bosonized Hamiltonian are of the form~\cite{OshikawaYamanakaAffleck_PhysRevLett.78.1984}
     \begin{multline}
      \delta H_{QFT} \propto \int dx \sin \left(\pi\varphi(x) + 2 k_F j  \mp \frac{2\pi \ell }{R}j \right).
     \end{multline}
      One of these terms for $\ell = 0,\ldots,R-1$ survives at long distances and can potentially gap out the system if short-distance oscillations are cancelled out, i.e., if
     \begin{equation}
          k_F = \pm \frac{\pi \ell }{R}. \label{eq:OYA_condition_kf}
     \end{equation}
     Using the relationship between $k_F$ and $\tan\Theta$, we recover the condition in \cref{eq:OYA_condition} which was formulated by Oshikawa, Yamanaka and Affleck (OYA) in the context of magnetic plateaus in spin chains~\cite{OshikawaYamanakaAffleck_PhysRevLett.78.1984}. The OYA condition of \cref{eq:OYA_condition,eq:OYA_condition_kf} is necessary but not sufficient for opening up a gap. Additionally, we also need the operator  $\sin \left(\pi\varphi\right)$ to be relevant. This occurs when $K<2$, which includes the free-fermion case ($K=1$), consistent with the above discussion.

     So far we focused on perturbations explicitly breaking translation symmetry. Interactions also introduce a new possibility of getting a gapped state by \emph{spontaneously} breaking translation symmetry. To illustrate this point, consider a translationally invariant perturbation of the form $ H_{1D} = \lambda \sum_j  \tau^z_{j} \tau^z_{j+1} \cdots \tau^z_{j+R-1}$. We discussed above how the $R=2$ case naturally emerges within our interface setup; it can be shown that 1D perturbations of this form for $R>2$ are naturally generated  by longer-range Ising couplings in 2D, but we will not discuss this further here.
     In the bosonized Hamiltonian, the most relevant term this perturbation introduces is
    \begin{equation}
        \delta H_{QFT} \propto  \int dx \sin \left(R\pi\varphi(x) + 2 k_F R j  \right) \, .
    \end{equation}
    This survives at long distances if $2 k_F R  = 0 \mod 2\pi$, i.e., if
    \begin{equation}
           k_F  = \pm \frac{\pi \ell }{R}  \, .
    \end{equation}
    This gives the same OYA condition as in \cref{eq:OYA_condition_kf,eq:OYA_condition}. However, in addition to this condition, we also need the operator $\sin(R\pi\varphi)$ to be relevant. This gives the more stringent condition $R^2K <2$, which rules out free fermions for any $R>1$ and requires strong repulsive interactions.

 In conclusion, we note that a horizontal modulation of the transverse field or of the Ising couplings with periodicity \textit{incommensurate} with the lattice (which can be viewed as the $R\to\infty$ limit of the above argument) generates a quasiperiodic hopping $g_j = g \cos(\alpha j + \beta)$ or potential $h_j = h \cos(\alpha j + \beta)$, leading to the pinning of interfaces with \textit{arbitrary} spatial orientation $\pi/4\le \Theta\le \pi/4$ via the mechanism of Anderson localization by pseudorandom quenched disorder~\cite{aubry1980analyticity}. This phenomenon has been proven robust to weak interactions~\cite{mastropietro2015localization}, thereby suggesting that the interface pinning phenomenon may be robust to arbitrarily weak additional bulk perturbations.

\section{Universal curvature-driven dynamics}

\label{sec_neq}

One of the most promising features of synthetic quantum matter is straightforward access to coherent out-of-equilibrium dynamics. The purely geometric degrees of freedom may be evolving far from equilibrium even when the bulk background is  frozen at zero temperature. Within the ``holographic'' description, the interface geometry is encoded in a 1D charge-density profile, thus creating a direct bridge between 1D hydrodynamic particle/spin transport --- a research topic that witnessed spectacular progress in the last decades~\cite{bertini2021finite} --- and 2D curvature-driven interface dynamics, which has only recently been started to be probed in quantum simulators~\cite{manovitz2025quantum}.

Let us start by discussing the validity of the ``holographic'' approach for non-equilibrium dynamics, considering first the perturbative regime $\lambda\ll J$. Within the setup of Sec.~\ref{sec_SW}, we consider time evolution driven by real-time control of the bulk Hamiltonian parameters $\vec\lambda=\vec\lambda(t)$,
exemplified by a
ramp $\vec\lambda(t)$ from $\vec\lambda_0 $ to $ \vec\lambda $ with speed $\dot{\lambda}$.
As observed in Sec.~\ref{sec_unveiling}, for speeds much smaller than the 2D bulk gap, i.e., $\dot{\lambda} \ll J^2$, by the adiabatic theorem the density of 2D bulk excitations (magnons) is parametrically suppressed, while the interface can still exhibit fast and complex far-from-equilibrium dynamics if $\dot \lambda \gg \lambda^2$.

Formally, rigorous prethermalization theorems guarantee that the number operator counting frustrated bonds $N_{\rm dw}=\sum_{\langle\mathbf{r},\mathbf{r}'\rangle}\frac {1-\sigma^z_\mathbf{r}\sigma^z_\mathbf{r'}}2$ is perturbatively close to a quasi-conserved operator $\widetilde{N}_{\rm dw}=U_{\rm SW} N_{\rm dw} U^\dagger_{\rm SW} = N_{\rm dw} + \mathcal{O}(\lambda/J)$ independent of energy, whose commutator with the Hamiltonian --- while non-vanishing --- is smaller than any power of the perturbation, $\lVert[\widetilde{N}_{\rm dw},H_{\rm 2D}]\rVert\le \Gamma\sim \exp(-J/\lambda)$~\cite{abanin2017rigorous,lin2017quasiparticle,lerose2020quasilocalized}. Thus, the expected systematic variations of $N_{\rm dw}$ during time evolution --- associated, e.g., with radiative emission of bulk excitations from the interface --- can only occur over a time scale $\Gamma^{-1}$ that grows exponentially large with the inverse perturbation strength. This shows that our effective 1D Hamiltonian accurately describes non-equilibrium interface dynamics over extremely  long  timescales, typically much beyond the coherence time of quantum simulators.

\begin{figure*}
    \centering
    \begin{subfigure}
    \centering
    \raisebox{9mm}{\includegraphics[width=0.27 \linewidth]{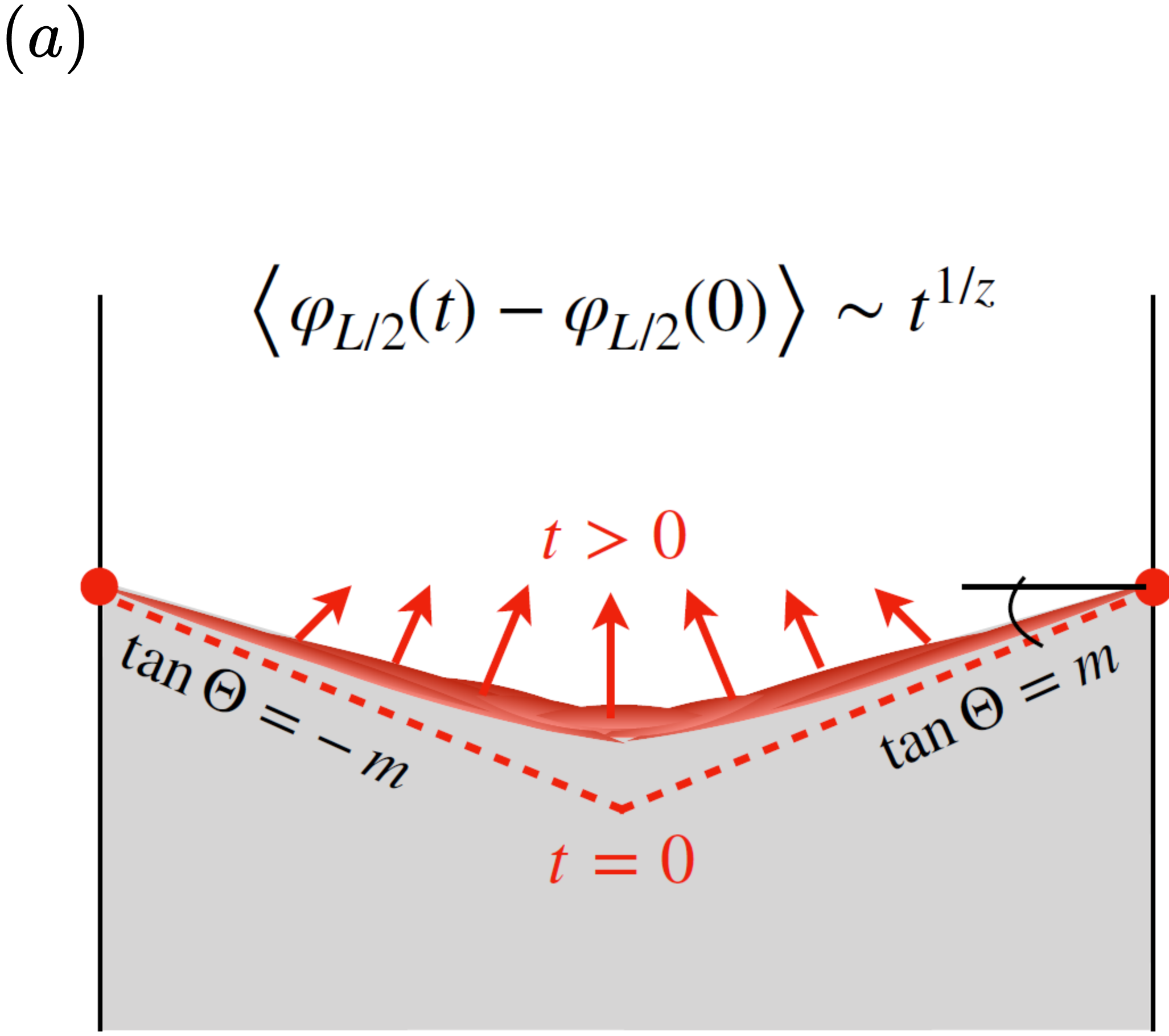}}
    \end{subfigure}
    \hspace{0.3cm}
    \begin{subfigure}
    \centering
\includegraphics[width=0.67\linewidth]{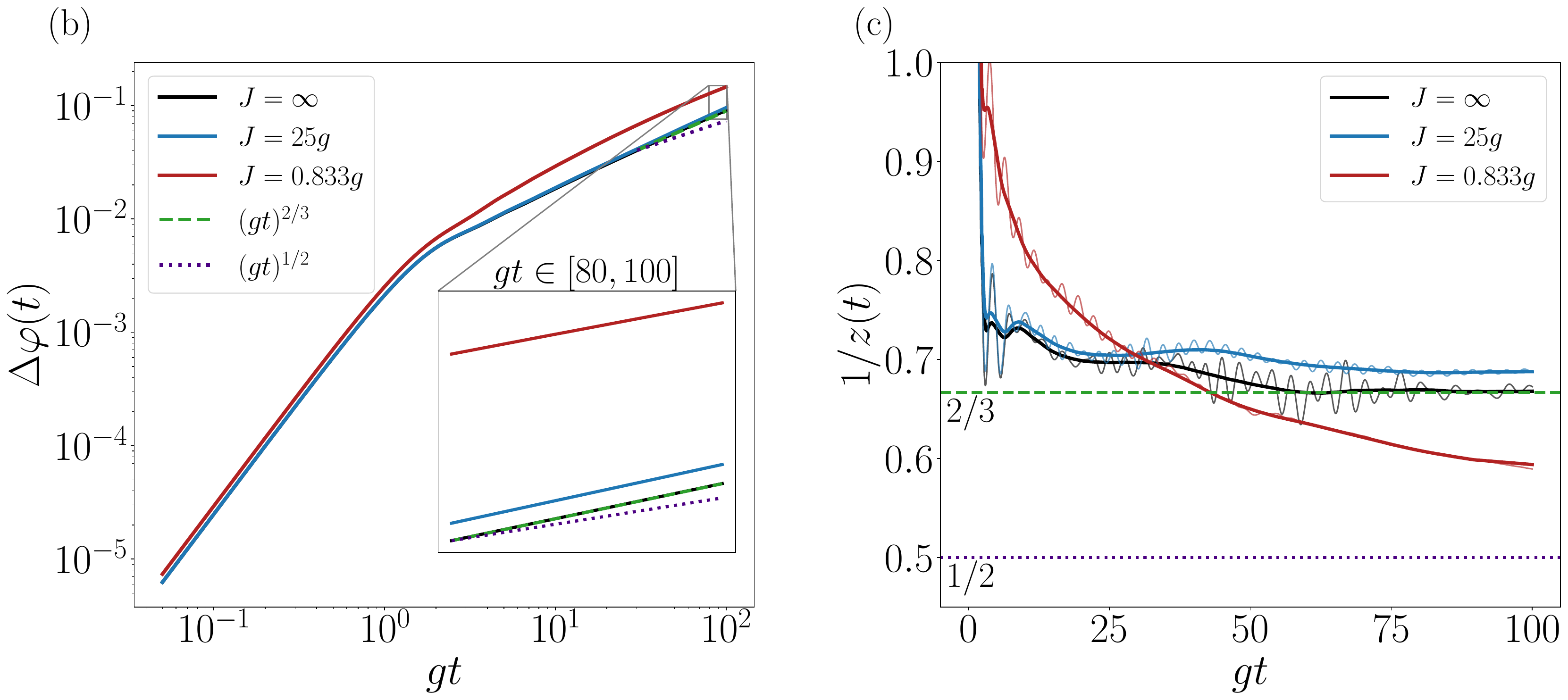}
\end{subfigure}
    \caption{
    \textbf{Curvature-driven quantum dynamics.} Panel (a): Illustration of curvature-driven evolution starting from a corner-like configuration at $t=0$.
    Panel (b): Evolution of the interface height variation $\Delta\varphi(t)=\varphi_{L/2}(t)-\varphi_{L/2}(0)$  from initial slopes  $\tan \Theta = \pm m= \pm  0.01$, computed from the effective 1D Hamiltonian dynamics~\eqref{eq_XXZ2o} for $J=\infty$ (black), $J=25g$ (blue) and $J=0.833g$ (red), corresponding to spin-transport in an unperturbed, weakly perturbed, and strongly perturbed $SU(2)$-symmetric Heisenberg chain of size $L=200$, respectively.
        Panel (c): Running dynamical exponent $1/z(t)=d\log\Delta \varphi/d\log t$ for the same three simulations, shown as  raw point-wise log-derivative (thin) and  smoothed sliding-window fit (thick).
        Dashed and dotted lines mark the KPZ ($2/3$) and diffusive ($1/2$) universal exponents.
    }
    \label{fig_neq}
\end{figure*}

As a case study, we will consider the relaxation dynamics of an interface configuration with a nonvanishing initial curvature~\cite{Alessio_PhysRevLett.129.120601,Alessio_PhysRevB.107.024306,manovitz2025quantum}. Within the ``holographic'' description, curvature --- i.e., a spatially varying {interface} slope in 2D --- maps to a spatially varying particle or magnetization density in 1D. In the simplest configuration, illustrated in Fig.~\ref{fig_neq}(a), the interface slope varies from a negative value attained at the left edge of the system to a positive value attained at the right edge (or viceversa). This configuration maps to the well-studied partitioning protocol for 1D  transport, with a net extensive particle number or magnetization imbalance between the left and right halves of the chain. The ensuing 1D density and current profiles, $m(x,t)=\langle \tau^z_j(t)\rangle$, $j(x,t)=\int^x dy \, \dot m(y,t)$ are generally found to follow spatiotemporal scaling laws,
\begin{equation}
    m(x,t) \underset{x,t\to\infty}{\thicksim} F(x/t^{1/z})
\end{equation}
characterized by a dynamical critical exponent $z$ and a scaling function $F$, where we set $x = j-L/2$ here. Through  ``holographic'' mapping, the interface curvature reads
\begin{equation}
    \kappa(x,t) = \frac{\partial^2_x \varphi(x,t)}{\left( 1+(\partial_x\varphi(x,t))^2\right)^{3/2}} = \frac{\partial_x m(x,t)}{\left( 1+m^2(x,t)\right)^{3/2}}.
\end{equation}
Hence, in the late-time regime, we find
\begin{equation}
    \label{eq_curvaturescaling}
    \kappa(x,t)\underset{x,t\to\infty}{\thicksim} t^{-1/z} F^\prime(x/t^{1/z}) \, .
\end{equation}
We conclude that \emph{the universality class of curvature-driven quantum dynamics of interfaces in 2D is directly linked to a corresponding universality class of non-equilibrium charge transport in 1D}.

This observation guides our understanding of experiments and allows us to make further predictions. For example, the experimental observation of a time-dependent radius of curvature satisfying the proportionality law $\dot R \propto 1/R$ in Ref.~\cite{manovitz2025quantum} --- analogous to classical phase-ordering dynamics~\cite{Bray1994PhaseOrdering} --- can be seen to be equivalent to our scaling law~\eqref{eq_curvaturescaling} with $z=2$, i.e., a \textit{universal diffusive scaling} in 1D. Indeed, the statement that the interface normal velocity $\dot \varphi$ is proportional to its average local curvature $\sim\partial_x^2 \varphi$ in 2D maps exactly to 1D Fick's law:
\begin{equation}
    \dot \varphi = - D \partial_x^2 \varphi \quad \Leftrightarrow \quad j = - D \partial_x m \, .
\end{equation}
 Considering the strong-perturbation regime of quantum antiferromagnets simulated in Ref.~\cite{manovitz2025quantum}, the experimental observation thus appears consistent with the  universality of diffusive transport in generic strongly interacting quantum spin chains.

Our analysis suggests that a broader range of universal curvature-driven relaxation laws can be observed in 2D synthetic quantum magnets as the holographic Hamiltonian $H_{\rm 1D}$ is tuned near integrability. For illustration, let us consider a 2D quantum Ising model with tunable interaction range and transverse field. (In Sec.~\ref{sec_qusim} below, we will discuss in detail the closely related model realized in the experiment~\cite{manovitz2025quantum}.) For short interaction range and weak field, curvature-driven interface dynamics map to spin-transport dynamics in an XXZ quantum spin chain~\eqref{eq_1DXXZ}, whose understanding has progressed tremendously in recent years~\cite{bertini2016transport,piroli2017transport,ljubotina2017spin,Bulchandani2018bethe,ljubotina2019kardar,collura2020domain,Misguich2019domain,scopa2022exact}. The transport universality class of this system is determined by the value of the anisotropy parameter $\Delta=2J'/g$ as follows:  diffusive  transport ($z=2$) in the gapped regime $\Delta>1$; superdiffusive transport ($z=3/2$) at the isotropic Heisenberg point $\Delta=1$, with a dynamical exponent compatible with the Kardar-Parisi-Zhang (KPZ) universality class, whose nature is still  debated~\cite{Gopalakrishnan2019kinetic,ilievski2021superuniversality,denardis2021stability,denardis2020Superdiffusion,wei2022quantum,andersen2024thermalization}; ballistic transport ($z=1$) in the gapless regime $|\Delta|<1$, with a Drude weight exhibiting a fractal dependence on $\Delta$ (an occurrence attributed to the structure of Bethe strings and conservation laws).\footnote{For the ``free-fermion'' case $\Delta=0$ the asymptotic scaling profile was computed exactly in Refs.~\cite{Alessio_PhysRevB.107.024306}.}
We thus predict
distinct transient scaling behaviors of curvature-driven interface dynamics as the transverse-field strength is increased in this setup.

To support this prediction, in  Fig.~\ref{fig_neq}(b,c) we report numerical results for realistic non-equilibrium curvature-driven dynamics, obtained by simulating 1D spin-transport dynamics with the full interface Hamiltonian
\begin{equation}
\label{eq_XXZ2o}
\begin{split}
    H_{\rm 1D} =  \sum_{j} & -\frac g2 \left(\tau^x_j \tau^x_{j+1} + \tau^y_j \tau^y_{j+1} \right) \\
    &
     + \left( J^\prime -\frac{3g^2}{16J}\right) \tau^z_j \tau^z_{j+1}  \\
    &  -\frac{g^2}{8J}( \tau^x_j  \tau^x_{j+2} +  \tau^y_j  \tau^y_{j+2})  \\
\end{split}
\end{equation}
including the leading higher-order perturbative corrections to Eq.~\ref{eq_1DXXZ} arising from the Schrieffer-Wolff transformation, which breaks both integrability and $SU(2)$ symmetry.\footnote{Denoting by $J^{(n-1)}$ the coupling strength between $n$-th-nearest neighboring spins, with $J\equiv J^{(0)}$ and $J'\equiv J^{(1)}$, it can be shown that $J^{(2)}$ does not contribute to $H_{\rm 1D}$, and the leading corrections from the tails are longer-range diagonal interactions proportional to $J^{(3)}$ and $J^{(4)}$.
For simplicity, we consider $J^{(n\ge3)}=0$ here. See also Sec.~\ref{sec_rydbergs}.}
For illustration, we focus on dynamics near the fine-tuned super-diffusive regime. We thus fix $J' \equiv g/2+3g^2/16J$, i.e., $\Delta=1$, and tune the relative strength of the transverse field $g/J$. To exactly realize the protocol sketched in Fig.~\ref{fig_neq}(a), we initialize the 1D spin chain in a magnetization-imbalanced initial state~$\rho(0)$, corresponding to an average interface profile with a central corner, and extract the running dynamical exponent $z(t)$ from the time-dependent transferred magnetization between the two halves of the chain, corresponding to the interface height relaxation at the center, shown in Fig.~\ref{fig_neq}(b).
For the simulations, following Ref.~\cite{ljubotina2017spin}, we take an infinite-temperature state of the form $\rho(0) = 2^{-L} \bigotimes_{j=1}^{L/2}(\mathbb{1} - m \tau^z_j) \otimes \bigotimes_{j=L/2+1}^{L}(\mathbb{1} + m \tau^z_j)$ with $0<m \ll 1$ and $L=200$.\footnote{Strictly speaking, an easier initial state to prepare in experiment is the corresponding ensemble of product states in the computational basis, i.e.,  $\rho^\prime(0)=U_{\rm SW} \rho(0) U^\dagger_{\rm SW} $. For small enough $|g/J|$ we expect no qualitative differences. }
 We evolve the traceless component $\delta\rho(t) = \rho(t) - \mathbb{1}/2^N$ as a vectorized matrix-product density operator using swap-gate TEBD with a singular-value truncation cutoff of $10^{-8}$ combined with a maximum bond-dimension cap $\chi=360$; the total truncation error, measured as the discarded weight relative to the full density matrix, stays well below $1\%$ for all curves shown.
Results in Fig.~\ref{fig_neq}(c) show that super-diffusive interface relaxation persists over a long transient for sufficiently small ratio $g/J$, making it accessible to quantum simulators with tunable-range spin-spin interactions. Interestingly, during this transient, the running value of the dynamical exponent $1/z(t)$ overshoots the exact KPZ value $2/3$ found for the $SU(2)$-symmetric Heisenberg chain by an amount that grows with the perturbation strength $|g/J|$. The super-diffusive transient is expected to eventually give way to universal diffusive relaxation, consistent with our simulations for larger values of~$g/J$.

As a central prediction of this work, in Sec.~\ref{sec_rydbergs} we will show that the quantum antiferromagnet experimentally realized in Ref.~\cite{manovitz2025quantum} can indeed be tuned through all the universal curvature-driven dynamical behaviors discussed here.

\section{``Holographic'' quantum simulation }
\label{sec_qusim}

\begin{figure}
\centering
\includegraphics[width=0.48\textwidth]{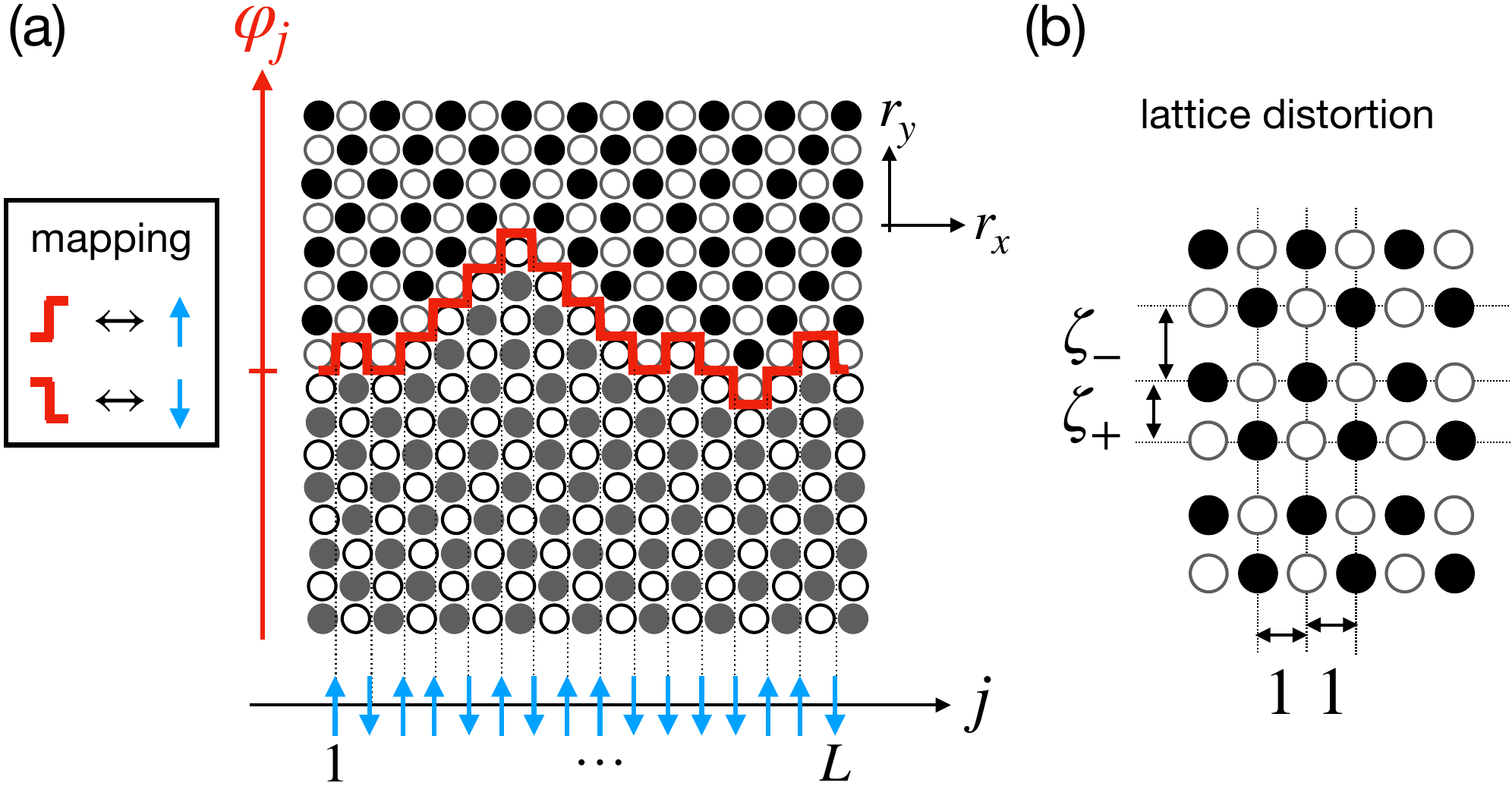}
\\
\vspace{0.4cm}
\includegraphics[width=0.92\linewidth]{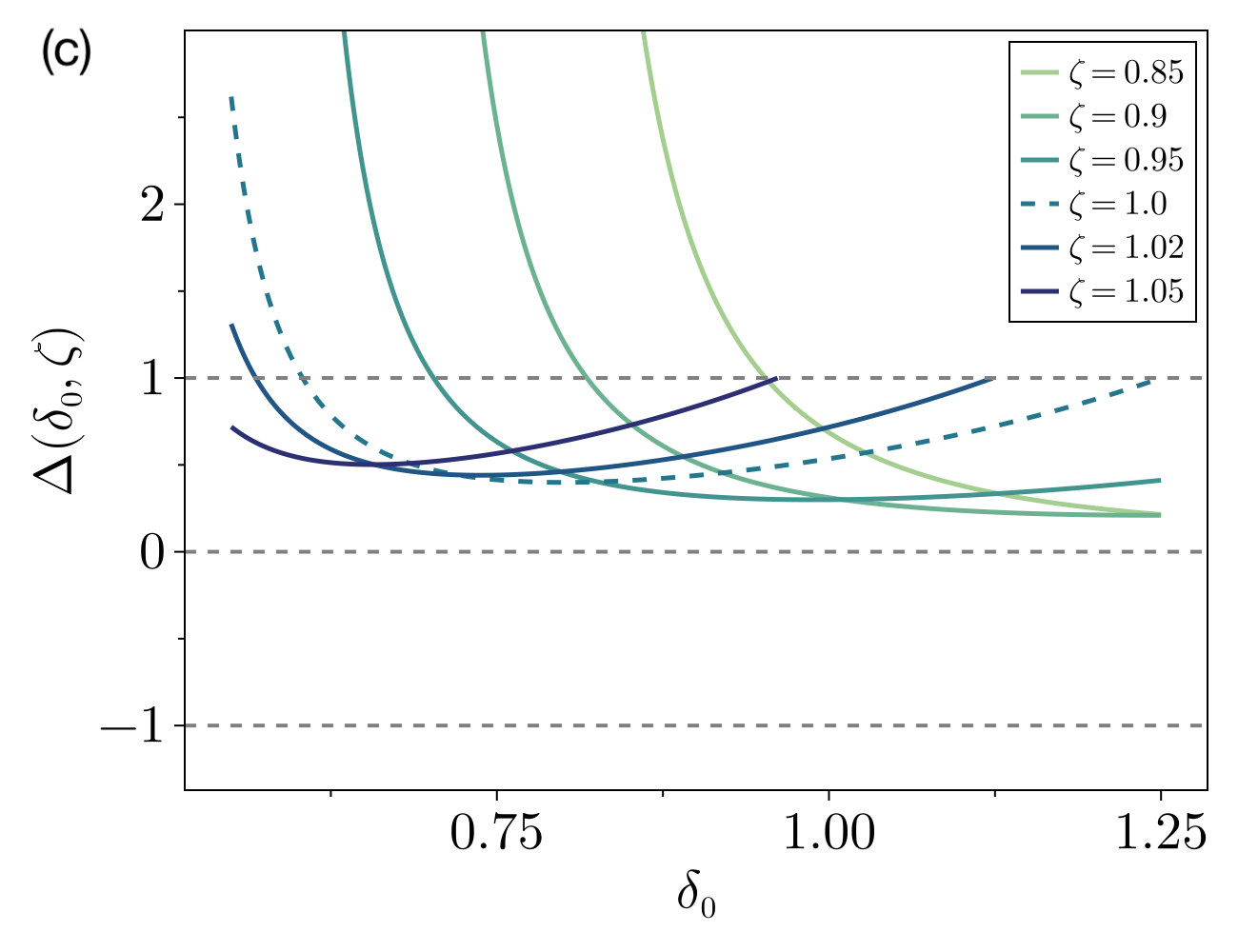}
\caption{
\mbox{\bf Neutral-atom interface engineering.}
Panel (a): Checkerboard (black-white) and anti-checkerboard (gray-white) ordered regions, composed of interleaved ground (empty dots) and Rydberg (full dots) atoms, separated by a low-energy interface. Such interface configurations can be mapped to a 1D spin chain similarly to Fig.~\ref{fig_mapping}.
Panel (b): Distorted rectangular array with vertical dilation
and staggering, giving rise to the effective interface Hamiltonian in Eq.~\eqref{eq_H1D_Rydberg}.
Panel (c): Anisotropy parameter $\Delta$ of the effective XXZ quantum spin chain, as a function of the experimental knobs $\delta_0=\Delta_0/V$ (detuning) and $\zeta=\zeta_+=\zeta_-$ (lattice anisotropy) in their allowed range. The dashed $\zeta=1$  curve corresponds to a conventional square array.
}
\label{fig_checkerboard}
\end{figure}

The approach and results of this work can be generalized to arbitrary 2D quantum systems spontaneously breaking a discrete symmetry, whenever boundary (or initial) conditions enforce an interface between domains of distinct degenerate ground states.
As a concrete illustration, here we will focus on synthetic quantum antiferromagnets realized in neutral atom arrays~\cite{ebadi2021quantum,scholl_quantum_2021,manovitz2025quantum}. In Sec.~\ref{sec_rydbergs} we will discuss how interfaces in this system can be tuned through the various static and dynamical  universality classes. Hence, in Sec.~\ref{sec_fcs} we will turn around our perspective and discuss the experimental advantages and viability of  ``holographically'' simulating an encoded 1D quantum many-body system.

\subsection{Interfaces in neutral-atom arrays}
\label{sec_rydbergs}

Neutral-atom hardware is at the heart of a rapidly developing technology for quantum simulation and computation.
This platform realizes fully configurable 2D arrays of atoms loaded in individual laser microtraps (optical tweezers) and coherently driven to a high-lying Rydberg state, resulting in strong van der Waals interactions.
The system is effectively described by the following many-body Hamiltonian,
\begin{equation}
\label{eq_Rydberg}
     H_{\rm{Ryd}} = \frac12 \sum_{ \mathbf{r}\neq\mathbf{r}^\prime} \frac {V}{|\mathbf{r}-\mathbf{r}^\prime|^6}  n_{\mathbf{r}} n_{\mathbf{r}^\prime} - \frac \Omega 2 \sum_{\mathbf{r}}  \sigma^x_{\mathbf{r}} -  \Delta_0 \sum_{\mathbf{r}}    n_{\mathbf{r}}  ,
\end{equation}
where the atom at site $\mathbf{r}=(r_x,r_y)$ is described by a two-level Hilbert space, spanned by its ground $|g_\mathbf{r}\rangle$ and Rydberg $|r_\mathbf{r}\rangle$ states, acted upon by $ \sigma^{x}_\mathbf{r}\equiv |g_\mathbf{r}\rangle\langle r_\mathbf{r}|+ |r_\mathbf{r}\rangle\langle g_\mathbf{r}|$ and $ n_{\mathbf{r}}\equiv|r_\mathbf{r}\rangle\langle r_\mathbf{r}|$. The parameters $\Omega$ and $\Delta_0$ represent the effective Rabi frequency and detuning, respectively, of the laser drive scheme, and the energy scale $V=\frac{C}{b^6}>0$ combines the relevant van der Waals coefficient~$C$ and a reference physical spacing~$b$ between the atoms.
For our study we will consider a square lattice and, to simplify the analysis, we will retain only nearest- and next-nearest-neighbor couplings, of strength $V$ and $V/8$, respectively; due to the rapid decay of the van der Waals tails, this is an excellent approximation frequently used in the literature.

The recent experiment in Ref.~\cite{manovitz2025quantum} demonstrated state preparation and observation of interface-dynamics between $\mathbb{Z}_2$-symmetry-breaking checkerboard and anti-checkerboard regions~\cite{Samajdar2020complex,ebadi2021quantum,scholl_quantum_2021}. For the classical Hamiltonian with $\Omega=0$, stability of these bulk order and domains requires $\sum_{nnn} V_{nnn} < \Delta_0 < \sum_{nn} V_{nn}$, where ${nn}$ and ${nnn}$ stand for the nearest and next-nearest neighbors.
Additionally, under the more restrictive condition
$\Delta_0 < V_{nn} + 2V_{nnn}$,
minimal-energy interfaces between checkerboard and anti-checkerboard regions are constituted by adjacent pairs of ground-state atoms. As shown in Fig.~\ref{fig_checkerboard}(a), since these defect lines naturally travel diagonally at $\pm 45^\circ$ across the lattice, the ``holographic'' mapping is obtained here without tilting the lattice.

The lattice geometry is fully configurable in experiments. In this Subsection it will be convenient for us to consider anisotropic rectangular arrays with an additional vertical staggering, given by coordinates $\mathbf{r}=(n_x,\zeta n_y+\varepsilon_\zeta (-1)^{n_y})$, $n_x,n_y\in\mathbb{Z}$ as shown in Fig.~\ref{fig_checkerboard}(b).
The relative vertical spacings are alternatively equal to $\zeta_\pm \equiv \zeta \pm 2\varepsilon_\zeta$ at even and odd rows, respectively; we will tune these spacings around unity (i.e., around the isotropic-lattice condition).
This simple lattice distortion will allow us to tune the effective 1D interactions and bond-dimerization parameters describing interface dynamics, realizing a generalization of the effective Hamiltonian in Eq.~\eqref{eq_1Dbonddimerized}.

Within our next-nearest-neighbor approximation all minimal-length interface configurations are classically degenerate.\footnote{It can be shown that the main effect of van der Waals tails is to introduce weak $ZZ$-type interactions of strength $\approx - 0.01 V$ in the induced 1D Hamiltonian. \label{footnote_tails}} We can then build a dictionary between experimentally relevant bulk perturbations to the Rydberg Hamiltonian~\eqref{eq_Rydberg} and 1D Hamiltonians, analogous to Table~\ref{tab_summaryIsing}.

The most important perturbation is given by
the transverse field, i.e., the Rabi term $\propto \Omega$ in Eq.~\eqref{eq_Rydberg}.  The effective 1D quantum dynamics is generated by second-order processes in $\Omega$.
Calculation shows that these processes induce a 1D bond-dimerized XXZ spin chain that generalizes Eq.~\eqref{eq_1Dbonddimerized}:
\begin{multline}
\label{eq_H1D_Rydberg}
    H_{1D} \; \propto \;    \sum_{j=1}^{L-1} \Big[ (1+(-1)^j\varepsilon_{XY})\,\big( \tau^x_j \tau^x_{j+1} + \tau^y_j \tau^y_{j+1}\big) \\+ \Delta (1+(-1)^j\varepsilon_{ZZ}) \, \tau^z_j \tau^z_{j+1} \Big] \, .
    \end{multline}
The explicit dependency of the dimensionless parameters $\Delta$, $\varepsilon_{XY}$, $\varepsilon_{ZZ}$  on the three dimensionless experimental knobs $\delta_0\equiv\Delta_0/V$, $\zeta_\pm$, as well as the overall proportionality constant, are reported in Appendix~\ref{app_engineering}.

In the absence of vertical staggering, i.e., for  $\zeta_+=\zeta_-\equiv\zeta$,
we have vanishing dimerization parameters $\varepsilon_{XY}=\varepsilon_{ZZ}=0$ and thus
we recover the standard XXZ spin chain with tunable anisotropy parameter $\Delta=\Delta(\delta_0,\zeta)$.
In Fig.~\ref{fig_checkerboard}(c) we plot the variation of the function $\Delta(\delta_0,\zeta)$ across the experimentally relevant parameter space. Our results show that the anisotropy parameter is tunable in the range $\Delta \in [0.187,+\infty)$.
For a standard isotropic atom array with $\zeta=1$, the minimum value of $\Delta$ is $0.399$, attained around $\delta_0=0.8$.
We have verified our analytical derivation of Eq.~\eqref{eq_H1D_Rydberg} by comparing numerical DMRG simulations of interface fluctuations in the full 2D system with charge fluctuations in the corresponding 1D chain, finding excellent agreement with the predicted value of $\Delta$ (see below). This result confirms that the quantum simulator can access all the curvature-driven universal behaviors anticipated in Sec.~\ref{sec_neq}, as $\Delta$ is tuned across $1$ in experiment.

We further note that since van der Waals tails beyond next-nearest-neighbors induce a small $\Omega$-independent additional ferromagnetic coupling $\approx -0.01 V \sum_j \tau^z_j\tau^z_{j+1}$ in the 1D Hamiltonian (cf. footnote~\ref{footnote_tails}), tuning $\Omega$ to sufficiently small values allows to vary~$\Delta$ below the positive minimum determined above, in principle all the way down to $\Delta=-\infty$ (at the price, however, of making the coherent dynamics slower and thus more susceptible to experimental noise accumulation).

In Appendix~\ref{app_engineering} we report the non-trivial iso-dimerization surface in knob space, defined by the equation $\varepsilon_{XY}(\delta_0,\zeta_+,\zeta_-)=\varepsilon_{ZZ}(\delta_0,\zeta_+,\zeta_-)$, along which the Rydberg-atom simulator realizes the exact bond-dimerized Hamiltonian in Eq.~\eqref{eq_1Dbonddimerized}.
We have numerically verified that, in agreement with field-theoretical bosonization predictions [cf. Eq.~\eqref{eq:Bosonization_XXZ_dimerization}], the experimentally accessible portion of the phase diagram of Eq.~\eqref{eq_H1D_Rydberg} covers \textit{all} the phases and phase transitions in Fig.~\ref{fig:combined_dimer}(a).

We finally note that a bulk perturbation of the form $- \sum_{ \mathbf{r},\mathbf{r'}} \gamma_{\mathbf{r},\mathbf{r'}}
\left(\sigma^x_{\mathbf{r}} \sigma^x_{\mathbf{r'}}+\sigma^y_{\mathbf{r}} \sigma^y_{\mathbf{r'}}\right)$ may be realized by encoding the atomic two-level system in two Rydberg states experiencing direct dipolar interactions --- giving rise to an XXZ-type Hamiltonian  with power-law decay $|\mathbf{r}-\mathbf{r}^\prime|^{-3}$ (see, e.g., \cite{ravets2015measurement}) --- or by  Floquet- and other engineering strategies~\cite{lerose2019prethermal,scholl2022microwave,ciavarella2023simulating,kunimi2025proposal}.
In these cases, interface quantum dynamics is driven by {first-order} resonant transitions in $\gamma$. Detailed analyses of these alternative (potentially more versatile) realizations of quantum-magnet interfaces are left for future investigations.

\subsection{Charge full counting statistics via local projective measurements }

\label{sec_fcs}

\begin{figure*}
    \centering
    \includegraphics[width=0.99\linewidth]{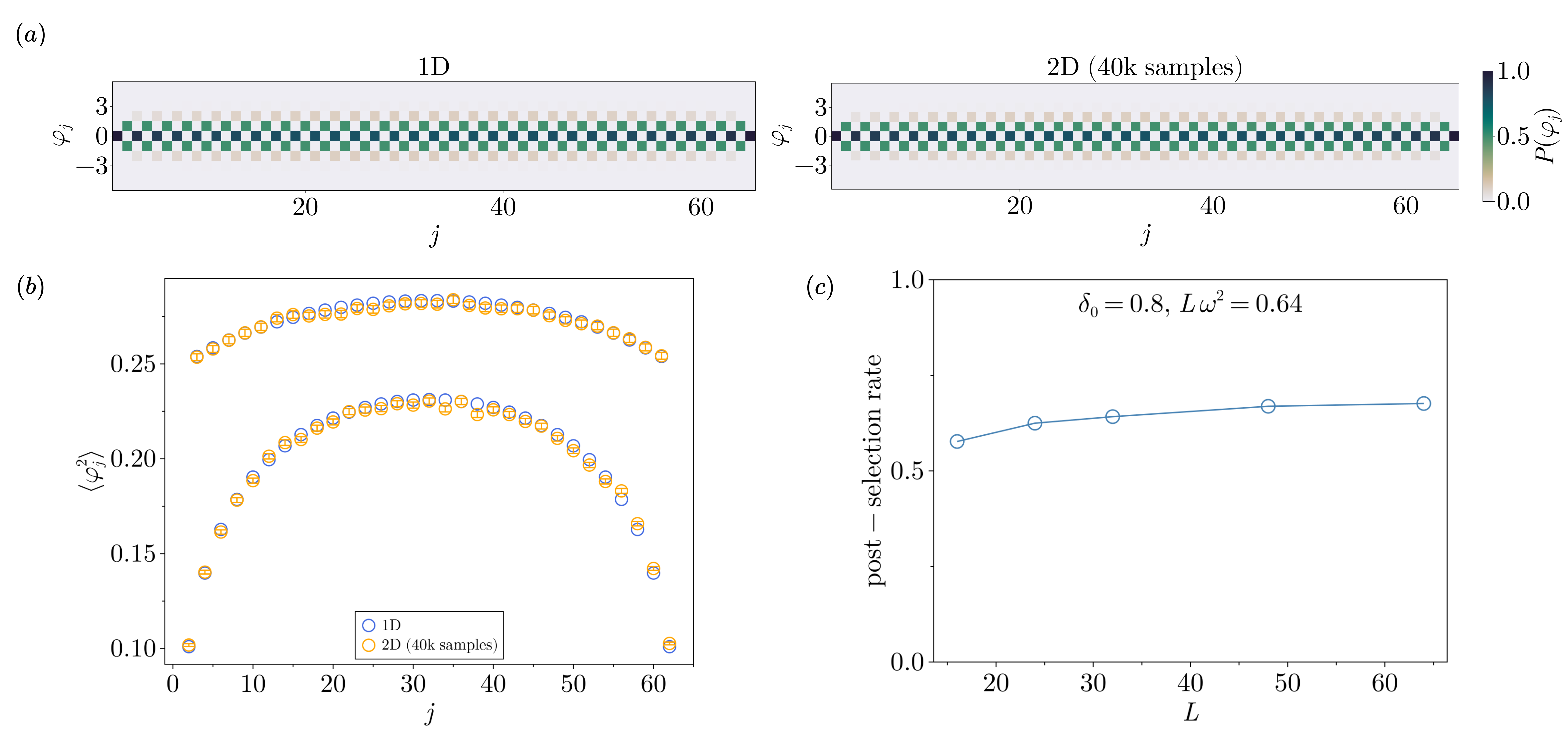}
    \caption{
    \mbox{\textbf{``Holographic'' reconstruction of full counting statistics  in an engineered neutral-atom array.}}
    Panel (a): Comparison between numerically computed charge full counting statistics in the ground state of a $L=64$ XXZ spin chain with $\Delta = 0.4$ (left panel), and the corresponding numerically simulated empirical  probability distribution of interface configurations in an engineered $(L+1) \times W=65\times12$ neutral-atom array with $\Delta_0=0.8 V$ and $\Omega=0.05 V$, from 40k post-selected ground-state samples (right panel). The average contrast between the two heat maps is around $1\%$ in probability. Panel (b): Comparison between site-resolved variances of charge fluctuations in the 1D XXZ ground state and transversal interface fluctuations in 2D (note the even-odd effect). Panel (c): Leakage of the ground-state wave function out of the 1D ``holographic'' subspace, measured by the post-selection rate of experimental snapshots with minimal-length domain wall, as a function of $L$ with \mbox{fixed $W=12$ and $L\omega^2=0.64$.}
    }
    \label{fig:combined_2D_dmrg}
\end{figure*}

Our dimensional-reduction framework for quantum interfaces may also be regarded in a reverse perspective as an enriched playground for quantum simulation of a vast family of 1D fermionic systems or spin chains using 2D (or rather quasi-1D) spin lattices.
This ``holographic'' quantum-simulation approach turns out to offer unique advantages, which we already briefly  mentioned above. For example, Ref.~\cite{lerose2024simulating} shows that $U(1)$ bosonic dynamical gauge fields can be naturally implemented via this route, overcoming a major challenge in the field of (analog) quantum simulation. Here we discuss a further unique advantage, that is,  enabling \textit{direct measurements of highly non-local 1D observables via simple local measurements in 2D}.

The most natural observable in our setup is given by the bosonic counting field $\varphi_{j}=2\sum_{l=1}^j n_l - j$ itself, corresponding to the interface height field (see Figs.~\ref{fig_mapping} and~\ref{fig_checkerboard}).
In the small-perturbation regime $\Omega\ll V$, imaging the atom located at~$\mathbf{r}$ --- i.e., measuring $n_\mathbf{r}$ --- unambiguously reveals whether the interface passes above or below~$\mathbf{r}$. This corresponds to the projective measurement $\Pi_{ \varphi_{n_x}< n_y} $.   As a result, measuring the defect operator on a vertical bond $n_\mathbf{r} n_{\mathbf{r}+\mathbf{e}_y}$ corresponds to the projective measurement $\Pi_{\varphi_{n_x}=n_y}$.
This shows that the 1D charge full counting statistics --- i.e., the probability distribution of the observable $\varphi_j$ --- is straightforwardly obtained through local measurements in 2D. We emphasize that this protocol implements a \textit{direct} measurement of the global charge contained in a large region \textit{without} measuring the entire charge distribution in the region, i.e., fully preserving quantum coherence within each subsystem charge sector and hence preserving symmetry-resolved entanglement~\cite{goldstein2018symmetry}.
Furthermore, since the measurement of $\varphi_j$ is through local operators $n_\mathbf{r} n_{\mathbf{r}+\mathbf{e}_y}$, it is only affected by experimental noise on a local portion of the array, in contrast with the extensive noise affecting the corresponding measurement in conventional 1D quantum simulator.

We have performed extensive numerical checks that the Rydberg quantum simulator operating under realistic experimental conditions allows indeed to reconstruct the full counting statistics of 1D charge from local projective measurements.
As an illustration, in Fig.~\ref{fig:combined_2D_dmrg} we report DMRG simulations of a 2D neutral-atom array described by the Hamiltonian $H_{Ryd}$ in Eq.~\eqref{eq_Rydberg} with $\zeta_{\pm}=1$ (conventional isotropic square lattice geometry), $\delta_0=\Delta_0/V=0.8$ and $\omega=\Omega/V=0.05$.
 For these parameter values, the ground-state wave function is a massive superposition of classical configurations of the array, ``holographically'' associated with the critical quantum fluctuations of a Luttinger liquid with $K \approx 0.7924$.
As shown in Eq.~\eqref{eq:LL_fluctuation_bosonization}, the scaling of height fluctuations is  only logarithmically diverging with $L$, which allows us to consider narrow quasi-1D ribbons in classical and quantum simulations, with negligible error.\footnote{While for simulations of rough phases one has $W\sim \log L$, for stiff phases one has an even more favorable scaling $W\sim \mathcal{O}(L^0)$.}
We simulate an array of size $(L+1)\times W=65 \times 12$, mapped  to a full $780$-site chain for optimal MPS encoding. We  enforce an interface by freezing the leftmost and rightmost column in symmetric classical configurations with a defect at half-height, and we always check convergence of observables with respect to the ribbon width $W$.\footnote{Due to the mentioned favorable scaling, a rapid convergence with $W$ is always attained. In practice, $W$ must only be chosen large enough that spurious states with the interface sticking to the top or bottom of the neutral-atom ribbon are sufficiently penalized in energy.  }

In Fig.~\ref{fig:combined_2D_dmrg}(a) we compare the empirical probability of finding the defect line passing through a given vertical bond upon imaging the array in the computational basis (right panel) to the probability distribution of the total charge across each bipartition in the corresponding 1D system (left panel), with no free parameters to fit. The empirical distribution is numerically obtained by sampling the MPS ground-state wave function in the computational basis and post-selecting on minimal-length domain-wall configurations (see below for discussion) --- thereby emulating the exact experimental procedure. The 1D full counting statistics is instead numerically computed as exact Fourier transform of the cumulant generating function (see, e.g., Ref.~\cite{Klich2003}).
The agreement appears excellent, even for the fairly large systems we simulated.
In Fig.~\ref{fig:combined_2D_dmrg}(b) we further compare the site-resolved variance of transversal interface fluctuations and 1D charge fluctuations.

The small discrepancy between the two results ($\left|P^{(1D)}(\varphi_j=q)-P^{(2D)}(\varphi_j=q)\right| \lesssim 1\%$ on average) is to be attributed to the truncation of the Schrieffer-Wolff  perturbative series to the lowest non-trivial perturbative order for both the Hamiltonian ($H_{\rm SW}$) and the state ($U_{\rm SW}$), cf. Sec.~\ref{sec_mapping}.
While  Hamiltonian approximation is a typical and well-understood feature of analog quantum simulation, neglecting the transformation of states requires some discussion following the caveats stated in \cref{sec_unveiling}. As pointed out before, an experiment typically performs measurements in the standard (non-Schrieffer-Wolff-transformed) computational basis. In fact, $U_{\rm SW}$ causes a leakage of the wave function out of the ``holographic'' subspace exactly mapped to 1D,   given by a weak perturbative admixture of minimal-length domain-wall configurations with ``non-holographic'' configurations featuring bulk defects (isolated magnons) and interface defects (overhangs), cf. Fig.~\ref{fig_mapping}(a). For sufficiently large perturbation $\omega$, this leakage may hinder the reconstruction of 1D observables.

As we argued in Sec.~\ref{sec_unveiling}, defects are expected to appear in  experimental snapshots of an  interface wave function
 with a probability per unit length\footnote{For simplicity, in this argument, we neglect the possible sublinear scaling of $W$ with $L$. } of order $\mathcal{O}(\omega^2)$, leading to an  infidelity between the actual (2D) and  ideal (1D ``holographic'') wave functions scaling as $1-\exp( -C L \omega^2)$ for some numerical constant $C=\mathcal{O}(1)$.
Post-selecting snapshots with minimal-length domain walls effectively implements a projection of the full 2D wave function back onto the defect-free ``holographic'' subspace where the target simulated 1D system lives. Although this projection does not fully restore perfect fidelity, the scaling of the empirical post-selection rate represents an indirect probe of the state infidelity.
In Fig.~\ref{fig:combined_2D_dmrg}(c) we report this rate as a function of the array's linear size $L$, with fixed $W$, and $\omega$ proportionally decreased as $L^{-1/2}$.
Data confirms that the fraction of discarded snapshots stays roughly constant, as suggested by the above argument.

Crucially, for small but finite (i.e., $L$-independent) perturbation $\omega$, the locality property of Schrieffer-Wolff generators $S^{(n)}$ as 2D operators (cf.~Sec.~\ref{sec_SW}) ensures that ``non-holographic'' interface defects are dilute and short-range correlated, making it possible to faithfully reconstruct 2D local observables with bounded error --- including $\varphi_j$ --- despite the exponential collapse of global fidelity and the accumulation of experimental noise~\cite{trivedi2025noise}.
By contrast, in a conventional 1D quantum simulator, the effects of local coherent and incoherent errors on charge full counting statistics $\varphi_j$ would necessarily be extensive in system size.
This observation, together with Fig.~\ref{fig:combined_2D_dmrg}(c), demonstrates the viability and utility of ``holographic''  quantum simulation.

\section{Outlook}

In this work we have shown that interfaces in 2D quantum magnets (equivalently, confining strings in dual lattice gauge theories) feature a rich
phase structure and non-equilibrium curvature-driven coherent evolution, ``holographically'' corresponding to ground-state charge fluctuations and non-equilibrium charge transport in suitable $U(1)$-symmetric 1D fermionic or quantum spin chains.

Crucially, contrary to most (if not all) previously studied quantum interfaces, generic points in the phase-diagram  portion we analyzed [cf. Eq.~\eqref{eq_ladder} and Fig.~\ref{fig:XXZ_ladder}] feature fluctuating quantum interfaces in 2D devoid of a 3D classical-statistical interpretation, generated by non-stoquastic Hamiltonians and thus inaccessible to efficient sampling schemes.

For non-equilibrium dynamics, we have recognized the scaling law experimentally observed in the recent Ref.~\cite{manovitz2025quantum} as a ``holographic'' manifestation of an underlying universal diffusive scaling in 1D, and we have determined parameter regimes where distinct dynamical scaling laws appear, including ``holographic'' manifestations of ballistic, super-diffusive,  and many-body localized behaviors. We have predicted how these behaviors can be experimentally observed with a programmable neutral-atom array.  We further  anticipate that more exotic curvature-driven relaxation phenomena may arise, for instance, from the interplay of distinct spin and charge transport behaviors in Fermi-Hubbard interfaces described by Eq.~\eqref{eq_1DHubbard}~\cite{fava2020spin}.

For experiments, we have proposed a new approach to quantum simulation of 1D quantum systems ---  which we dubbed ``holographic'' quantum simulation --- through encoding in a 2D (in practice  quasi-1D) quantum-magnet interface.
A central advantage of this approach is to  enable manipulation and observation of non-local information using simple local operations.
As a key application, we have proposed and numerically demonstrated a protocol for local projective measurement of the total charge contained in an arbitrarily large region: Measurements at the boundaries of the region project the state onto a definite partition of the total charge between the interior and the exterior \textit{without} affecting the coherent superposition of charge distributions in the two subsystems.
This possibility is beyond the reach of conventional 1D quantum simulators, where, furthermore, measuring charge full counting statistics is typically challenging in itself due to extensive accumulation of experimental noise~\cite{wienand2024emergence,Gopalakrishnan2024distinct,rosenberg2024dynamics,samajdar2024quantum,joshi2025measuring}. Our protocol thus opens the door to measurements of symmetry-resolved observables and entanglement~\cite{goldstein2018symmetry} as well as global projected ensembles and, potentially, new associated  measurement-induced phenomena~\cite{Milekhin2025observableprojected}.
In addition, it can  be shown that arbitrary measurements and feedback unitary operations at a position $j$ \textit{conditioned on} the total charge $\varphi_{j}=n_y$ can be straightforwardly implemented with purely local operations in 2D.

The most important theoretical question left open by our work is about the fate of ground-state phases and non-equilibrium universality classes in the fully non-perturbative region, including the neighborhood of the bulk critical point, where magnetic order eventually melts (or, in the dual gauge-theory picture, string tension vanishes and charges deconfine). Experience from ``effective string theory''~\cite{luscher1981thick,Caselle_2005,aharony2013effective}  suggests that all the interface phases we discussed eventually roughen and transition to a $c=1$ (1+1)D conformal field theory as a bulk critical point is approached, with largely universal finite-size corrections compatible with a purely geometric action. Details of the onset of this potential ``super-universality'' at the bulk critical point remain to be clarified, as well as its potential counterpart in real-time dynamics.

A related intriguing future prospect is the study of dynamical processes involving  the (gapped or gapless) fermionic/spin interface excitations across the various phases on the one hand, and the gapped bulk excitations --- i.e., magnons or glueballs in quantum-magnet or lattice-gauge-theory language, respectively --- on the other hand. These processes may, for example, generate an effective \textit{friction} in curvature-driven dynamics, or furnish a  cooling channel for experimentally preparing low-temperature correlated quantum states within ``holographic'' quantum simulation. This question inevitably appears as the bulk critical point is approached, effectively making the interface a lower-dimensional open quantum system.

Lastly, on a more technical level, this work opens several natural lines of investigation, which we  plan to pursue in the future. In one of these, we plan to extend our exact-bosonization framework to interfaces/strings wrapped around a long  cylinder, whose center-of-mass position --- i.e., the zero-mode of the non-compact height field --- exhibits  quantum dynamics beyond the scope of this work.
 In another, it would be interesting to further analyze and test the predicted robustness of Anderson-localized interface ground states. More broadly, it would be interesting to construct models where the \textit{geometric} interface degrees of freedom discussed in this work couple to \textit{dynamical} degrees of freedom of the kind often appearing in (fixed-)boundary quantum field theories, potentially generating an even richer quantum phase diagram and non-equilibrium relaxation behavior.

\begin{acknowledgments}
We thank Federico Balducci, Michele Caselle, John Chalker, Fabian Essler, Thierry Giamarchi, Joseph Indekeu, Tony Jin, Sara Murciano, Giuseppe Santoro, and Romain Vasseur for interesting discussions and suggestions.

We acknowledge funding through a Leverhulme-Peierls Fellowship at the University of Oxford (A.L.),  the FWO-FNRS EOS Research Project G0H1122N EOS 40007526 CHEQS (J.S.R.), the European Research Council (ERC) under
the EU Horizon 2020 Research and Innovation Programme during the early stages of this work (Grant Agreement No. 804213-TMCS;
A.P., S.A.P.), and a UKRI Frontier Research Grant
EP/Z002419/1 (S.A.P.). 

AI tools (OpenAI's ChatGPT and Codex) were used for minor assistance with drafting and polishing portions of the manuscript, proofreading,  source-file and bibliography cleanup, and polishing numerical code scripts. The authors reviewed and verified all outputs. All scientific content, analyses, interpretations, and conclusions were developed and verified by the authors, who take full responsibility for the manuscript.
\end{acknowledgments}

\appendix
\section{Interface variables and bosonization}
\label{app_soft}

In this Appendix, for the benefit of readers familiar with field-theoretical bosonization, we provide a more in-depth discussion of the exact bosonization in Sec.~\ref{sec_phases} and its subtleties.

\subsection{From the compact boson to the physical height}

There are two complementary descriptions of the interface.  Kinematically, each interface configuration is a path in two
dimensions and is therefore specified by its transverse coordinate, the
integer-valued height $\varphi_j$.  Dynamically, the elementary {\it slopes} of this
path are spin-$1/2$ variables,
\begin{equation}
 \varphi_j-\varphi_0=\sum_{l=1}^{j}\tau_l^z,
 \qquad
 \varphi_j-\varphi_{j-1}=\tau_j^z=\pm1 .
 \label{eq:height_slope_app}
\end{equation}
The Hamiltonian that controls the fluctuations of these slopes is a $U(1)$-symmetric spin-chain Hamiltonian.
Thus the configurations are those of a height field, whereas their quantum
dynamics can be analyzed by bosonizing a spin chain.

For definiteness, let us consider the archetypal XXZ Hamiltonian~\eqref{eq_1DXXZ}. Standard bosonization of this spin chain introduces a compact field
$\tilde\varphi\equiv\tilde\varphi+2$.  On an interval of length $L$, it is
useful to separate its zero mode, winding, and oscillator modes.  The
corresponding decomposition of the real-valued interface field may be
written as
\begin{multline}
 \varphi(x)
 =
 \overline{\varphi}
 +\frac{M}{L}\left(x-\frac{L}{2}\right)
 +\delta\varphi(x),
 \nonumber\\
 \int_0^L dx\,\delta\varphi(x)=0,
 \qquad
 \delta\varphi(L)=\delta\varphi(0).
 \label{eq:height_mode_decomposition_app}
\end{multline}
For the full system, the total magnetization is the integrated interface
slope.
\begin{equation}
 M =\int_0^L dx\,\partial_x\varphi(x) =\varphi(L)-\varphi(0) =
 2w
 \label{eq:height_winding_app}
\end{equation}
 and $w\in\mathbb Z$ is the winding of the compact
boson.  This gives a direct geometrical meaning to each part of the usual
compact-boson mode expansion:
\begin{itemize}
 \item $\overline{\varphi}=L^{-1}\int_0^L dx\,\varphi(x)$ is the absolute
 transverse center of mass of the interface.  Spin bosonization retains this
 zero mode only modulo $2$; the interface specifies its absolute lift/ unwinding.
 \item The winding is the net change in height.  Equivalently,
 $M/L=2w/L$ is the average slope, or tilt, of the interface and is the
 magnetization density of the XXZ chain.
 \item The oscillator modes describe shape fluctuations within a fixed
 zero-mode and winding sector.  More precisely, their gradients give the
 smooth part of the local slope fluctuations,
 \begin{align}
  \partial_x\varphi(x)
  &=\frac{M}{L}+\partial_x\delta\varphi(x),
  \nonumber\\
  \tau_j^z-\frac{M}{L}
  &\leftrightarrow
  \left.\partial_x\delta\varphi(x)\right|_{x=j}.
  \label{eq:height_gradient_dictionary_app}
 \end{align}
\end{itemize}

Equivalently, on any subinterval one may unwind the compact field according to
\begin{align}
 \varphi(x)-\varphi(0)
 &=
 2w(0,x)
 +\tilde\varphi_{\rm rep}(x)-\tilde\varphi_{\rm rep}(0),
 \label{eq:height_unwrapping_app}
\end{align}
where $\tilde\varphi_{\rm rep}$ is a representative modulo $2$ and $w(0,x)\in\mathbb Z$
records the number of compactification intervals crossed by the chosen
continuous unwinding.  On an open subinterval this winding does not represent any conserved charges. Only over the full system, with compatible
endpoint conditions, the winding $w(0,L)$ is related to the total magenetization.
\cref{eq:height_slope_app} is the exact lattice counterpart of
this statement: once one endpoint height is given, integrating the slopes
determines every $\varphi_j$ without ambiguity.  The zero-mode lift, the total
winding, and the local gradients therefore reconstruct a unique physical
interface configuration.

This also explains why the two uses of $\varphi$ in the main text are
consistent.  Local spin operators contain only derivatives of the field and
periodic functions such as $\exp(i\pi\ell\tilde\varphi)$, and are consequently
insensitive to $\tilde\varphi\to\tilde\varphi+2$.  The interface height is
instead the accumulated spin, and hence a non-compact observable:
\begin{equation}
 \varphi_j-\varphi_0=\sum_{l=1}^{j}\tau_l^z . \label{appeq:varphi_sigmaz}
\end{equation}
It is exactly the accumulated magnetization in $[1,j]$; equivalently,
the fermion number in that interval is
\begin{equation}
 N_j=\sum_{l=1}^j c_l^\dagger c_l
 =\frac{j+\varphi_j-\varphi_0}{2}.
\end{equation}
Thus the distribution of the height is equivalent, by change
of variables, to the charge full counting statistics.  The interface
representation retains the integer winding that local spin operators discard.
For pinned endpoints the boundary condition fixes the reference height and
the total winding.  For an interface on a cylinder, by contrast, its absolute
center of mass is an additional degree of freedom with no counterpart in the
slope-only spin description.

\subsection{Softening the height constraint}

A complementary way to obtain the continuum dynamics is to work directly
with the height field.  We relax the integer-valued height to a real variable
$\varphi_j\in\mathbb{R}$ and introduce its conjugate momentum $\Pi_j$.  The
Gaussian height Hamiltonian is
\begin{align}
 H_{\rm height}
 &=
 \sum_j\Bigg[
 \frac{2vK}{\pi}\Pi_j^2
 +\frac{\pi v}{8K}
 \left(\varphi_{j+1}-\varphi_j\right)^2
 \Bigg]+\ldots ,
 \label{eq:height_gaussian_app}
\end{align}
where $[\varphi_j,\Pi_l]=i\delta_{jl}$.  The first term is the kinetic energy
of the height field, while the second is the elastic energy associated with
spatial variations of the interface.  In the continuum limit,
$\varphi_{j+1}-\varphi_j\rightarrow\partial_x\varphi$, and
Eq.~\eqref{eq:height_gaussian_app} gives the Gaussian Luttinger-liquid
Hamiltonian.

The discreteness of the microscopic height is restored softly by adding a
periodic locking potential,

\begin{equation}
 H_{\rm soft}
 =
 H_{\rm height}
 -\Gamma\sum_j\cos(2\pi\varphi_j)+\ldots
 \label{eq:height_soft_app}
\end{equation}
For finite $\Gamma$, non-integer heights are allowed but cost energy; as
$\Gamma$ is increased, the field becomes localized near
$\varphi_j\in\mathbb{Z}$.  This is the soft-height implementation of the
microscopic height discreteness.

The locking potential has a minimum at every integer, but these local minima
do not give infinitely many ground states in the geometry of the main text.
The endpoints of the interface are pinned, and we choose the height origin so
that
\begin{equation}
	\varphi_0=\varphi_L=0,
\end{equation}
so the elastic and boundary energies select the flat configuration
$\langle\varphi\rangle=0$.  Here ``flat'' means flat on average;
individual microscopic paths still fluctuate. We will return to this in the next subsection.  A configuration centered on
another integer either translates the whole interface, which is incompatible
with the pinned endpoints, or introduces boundary kinks.  The boundary
condition therefore fixes the integer unwinding selected by the ground state.

\begin{figure*}[t!]
        \centering
                        \includegraphics[width=0.4\linewidth]{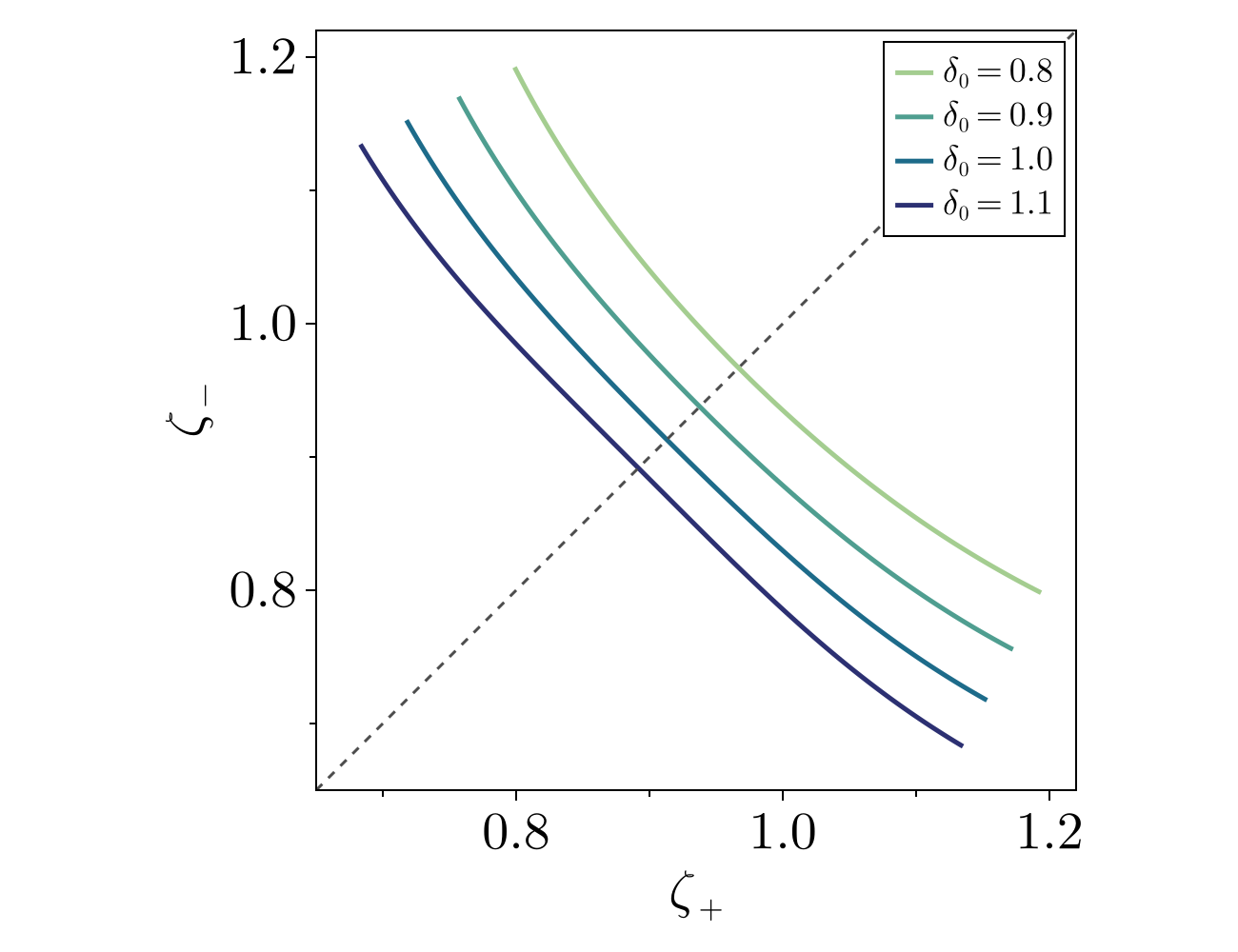}
        \includegraphics[width=0.4\linewidth]{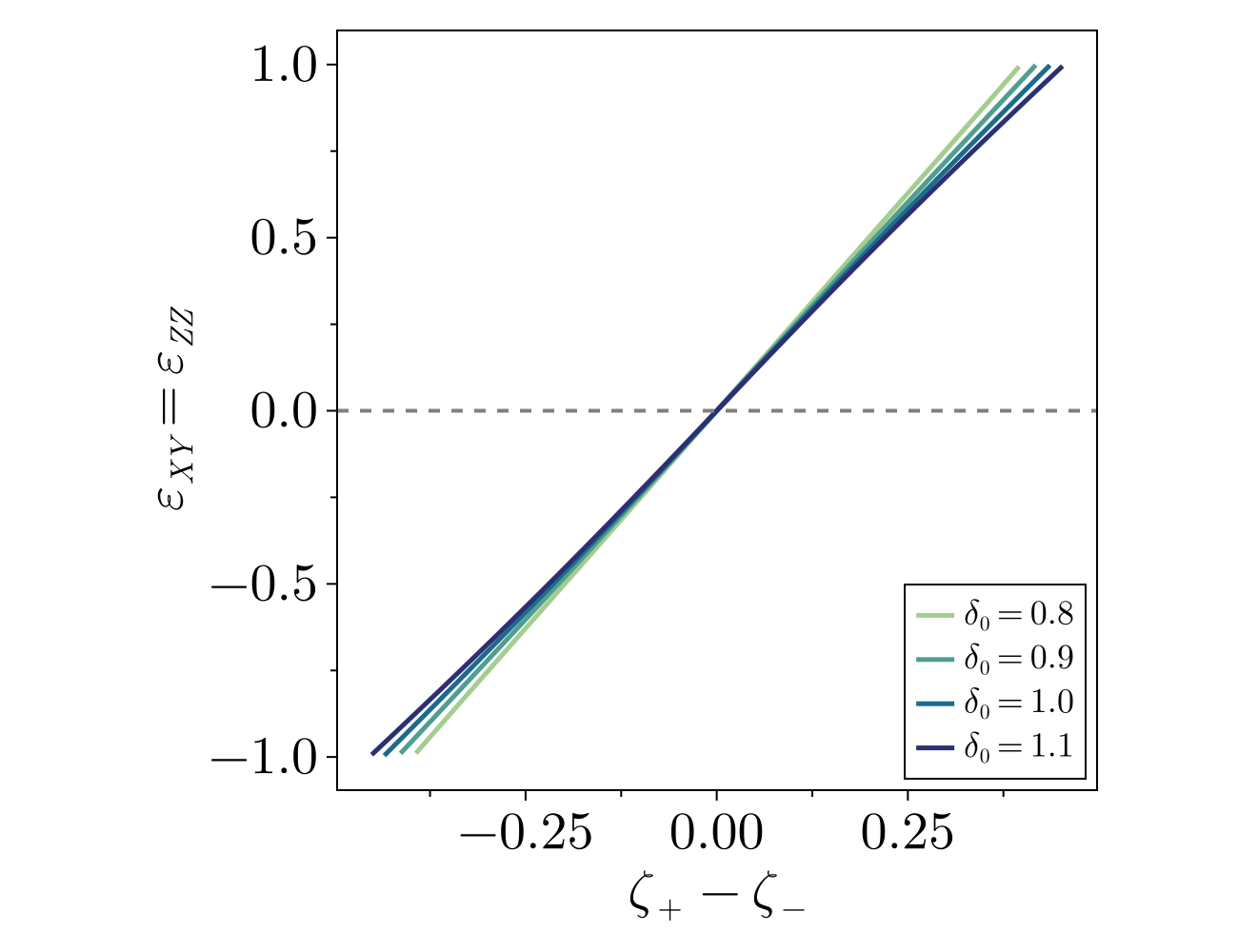}
        \caption{Iso-dimerization curves $\varepsilon_{XY}(\delta_0,\zeta_\pm)=\varepsilon_{XY}(\delta_0,\zeta_\pm)$ for various fixed values of the detuning $\delta_0$ in the allowed range (top panel) and the values of the common dimerization parameter $\varepsilon$ on these curves (bottom panel). The dashed diagonal line highlights the trivial iso-dimerization curve $\varepsilon_{XY}=\varepsilon_{ZZ}=0$.
        }
    \label{fig:iso-dimerization}
\end{figure*}

\subsection{The built-in LSM constraint and ground-state counting}

The periodic locking term has a more specific physical meaning here than in a
generic soft-height or Restricted-Solid-On-Solid (RSOS) model.  The slopes of this interface are the spins
of a zero-magnetization spin-$1/2$ chain, so the allowed locking operators inherit
the Lieb--Schultz--Mattis (LSM) constraint.  From \cref{appeq:varphi_sigmaz}, we have
\begin{equation}
 \varphi_j-\varphi_0\equiv j\pmod 2, \label{appeq:even_odd}
\end{equation}
and one-site translation acts on the long-wavelength field as
\begin{equation}
 T_1:\varphi\longmapsto\varphi+1 .
 \label{eq:height_translation_app}
\end{equation}
Consequently $\cos(\pi\varphi)$ is odd under one-site translation,
whereas $\cos(2\pi\varphi)$ is allowed.  Thus the same term that
implements integer locking in the height description is, in the spin
language, the VBS perturbation allowed by the LSM constraint.  For
$\Gamma>0$, $-\cos(2\pi\varphi)$ has a minimum for each integer valued $\varphi$. With the compactification however, $\tilde \varphi$ has two inequivalent minima, $\tilde\varphi=0$ and $\tilde\varphi=1$, which are
exchanged by translation.  When relevant, it therefore produces the two
translation-related vacua of the spin-chain VBS phase, as discussed in the
main text.  (The opposite sign instead selects the two half-integer minima
associated with N\'eel order)~\footnote{The VBS nature is also visible from the microscopic height fields $\varphi_j$ which are constrained to take odd (even) integer values on odd (even) lattice sites from \cref{appeq:even_odd}. So, deep in the VBS phase, flatness $\langle\varphi \rangle = 0$ is realized by genuine flatness on even sites  and a fluctuating height configuration on odd sites with a vanishing average, reminiscent of the resonating dimers of a VBS.}.
This is the bosonized form of the
LSM/OYA constraint at zero magnetization~\cite{OshikawaYamanakaAffleck_PhysRevLett.78.1984}. The same conclusion is implicit in the surface--spin-chain correspondence of
den Nijs and Rommelse~\cite{dennijs1989preroughening}: the disordered-flat
surface maps onto a dimerized VBS phase of a half-integer spin chain. In modern
LSM language, this gapped phase evades the obstruction to a unique symmetric
ground state by spontaneously breaking one-site translation and producing two
translation-related vacua, precisely as in the compact description above.
For the non-compact field, the pinned boundary fixes
the additive ambiguity in lifting the compact field, but also selects one of the
translation-broken sectors, so that the selected interface ground state is
the flat vacuum $\langle\varphi\rangle=0$.

The same distinction resolves the apparent mismatch in ground-state counting in other phases. Values separated by two represent the same compact spin field but different
non-compact lifts; they are nevertheless degenerate only if the boundary and
kink energies make them so.  For the explicitly dimerized theory,
$-\cos(\pi\varphi)$ selects the trivial-dimer vacuum
$\langle\varphi\rangle=0$, while $+\cos(\pi\varphi)$ selects the two nearest
odd lifts $\langle\varphi\rangle=\pm1$.  The latter carry opposite
half-kinks at the boundaries and give the two topological-dimer ground
states in the zero-magnetization sector.  More distant even or odd lifts
contain additional kinks and are excited states.  Finally, the
boundary-selected VBS state and the trivial dimer can both have the same flat
height profile; they remain distinguishable through their translation
properties and their lowest-lying kink and boundary excitations.\\

\section{Neutral-atom array interface engineering}
\label{app_engineering}

Here we report the result of the derivation of Eq.~\eqref{eq_H1D_Rydberg} from a second-order Schrieffer-Wolff transformation.

The dimensionless parameters $\Delta$, $\varepsilon_{XY}$, $\varepsilon_{ZZ}$  depend on the three dimensionless experimental knobs $\delta_0\equiv\Delta_0/V$, $\zeta_\pm$ as follows:

\begin{widetext}
    \begin{equation}
        \varepsilon_{XY} = \frac {f_+-f_-} {f_++f_-} \qquad
        \varepsilon_{ZZ} = \frac {g_+-g_-} {g_++g_-} \qquad
        \Delta = \frac {g_++g_-} {f_++f_-}
    \end{equation}
    where,
    \begin{equation}
        \begin{split}
        f_{\pm} = & + \frac{1}{\delta_0-\frac{1}{\zeta_{\mp}^6} + \frac{2}{(\zeta_\pm^2 + 1)^3} }
                + \frac{1}{  \frac{2}{(\zeta_\pm^2 + 1)^3}-\delta_0 }
        \\
        g_{\pm} =& + \frac{1}{\frac{1}{\zeta_{\mp}^6} + \frac{2}{(\zeta_\pm^2 + 1)^3} - \delta_0} +
                    \frac{1}{2 + \frac{1}{\zeta_{\mp}^6} - \delta_0} +
                    \frac{1}{\delta_0 -  \frac{2}{(\zeta_\pm^2 + 1)^3}} +
                    \frac{1}{\delta_0 -  \frac{2}{(\zeta_\pm^2 + 1)^3} - \frac{2}{(\zeta_{\mp}^2 + 1)^3}} \\
                    & - \frac{2}{1 + \frac{1}{\zeta_{\mp}^6} + \frac{1}{(\zeta_\pm^2 + 1)^3} - \delta_0} -
                    \frac{2}{\delta_0 -  \frac{2}{(\zeta_\pm^2 + 1)^3} - \frac{1}{(\zeta_{\mp}^2 + 1)^3}} \, ;
        \end{split}
    \end{equation}
\end{widetext}
the overall proportionality constant in $H_{1D}$ is $(f_++f_-)\Omega^2/{2V}$.

In Fig.~\ref{fig_checkerboard}(c) we visualize the dependency of $\Delta$ on the dimensionless detuning $\delta_0$ and uniform lattice distortion $\zeta_+=\zeta_-=\zeta$, while in Fig.~\ref{fig:iso-dimerization} we visualize the iso-dimerization curves $\varepsilon_{XY}(\delta_0,\zeta_\pm)=\varepsilon_{XY}(\delta_0,\zeta_\pm)$ corresponding to the Hamiltonian~\ref{eq_1Dbonddimerized}.

Corrections to the effective interface Hamiltonian in Eq.~\eqref{eq_H1D_Rydberg} scale as $\mathcal{O}\left(\Omega^4/V^3\right)$.

\bibliography{biblio}

\end{document}